\pdfoutput=1

\newcommand*{\ATLASLATEXPATH}{}
\documentclass[cernpreprint, atlasdraft=false, texlive=2016, UKenglish]{\ATLASLATEXPATH atlasdoc}
\usepackage[biblatex=false]{\ATLASLATEXPATH atlaspackage}

\usepackage{\ATLASLATEXPATH atlasbiblatex}

\usepackage{\ATLASLATEXPATH atlasphysics}

\AtlasTitle{Correlated long-range mixed-harmonic fluctuations measured in $pp$, $p$+Pb and low-multiplicity Pb+Pb collisions with the ATLAS detector}

\AtlasAbstract{%
Correlations of two flow harmonics $v_n$ and $v_m$ via three- and four-particle cumulants are measured in 13 TeV $pp$, 5.02 TeV $p$+Pb, and 2.76 TeV peripheral Pb+Pb collisions with the ATLAS detector at the LHC. The goal is to understand the multi-particle nature of the long-range collective phenomenon in these collision systems. The large non-flow background from dijet production present in the standard cumulant method is suppressed using a method of subevent cumulants involving two, three and four subevents separated in pseudorapidity. The results show a negative correlation between $v_2$ and $v_3$ and a positive correlation between $v_2$ and $v_4$ for all collision systems and over the full multiplicity range. However, the magnitudes of the correlations are found to depend on the event multiplicity, the choice of transverse momentum range and collision system. The relative correlation strength, obtained by normalisation of the cumulants with the $\langle v_n^2\rangle$ from a two-particle correlation analysis, is similar in the three collision systems and depends weakly on the event multiplicity and transverse momentum. These results based on the subevent methods provide strong evidence of a similar long-range multi-particle collectivity in $pp$, $p$+Pb and peripheral Pb+Pb collisions.
}

\author{The ATLAS Collaboration}

\PreprintIdNumber{CERN-EP-2018-151}

\AtlasDate{\today}
\AtlasJournalRef{Phys.\ Lett.\ B. 789 (2019) 444}
\AtlasDOI{10.1016/j.physletb.2018.11.065}

\usepackage{float} 
\usepackage{amssymb}
\usepackage{url}
\usepackage{harpoon}
\usepackage{amsmath}
\usepackage{MnSymbol}
\usepackage{multirow}
\usepackage{bm}
\usepackage{cite}

\newcommand{\lr}[1]{\left\langle #1\right\rangle}
\newcommand{\llrr}[1]{\left\llangle #1\right\rrangle}
\newcommand{\pPb}{\mbox{$p$+Pb}}

\newcommand{\pA}{\mbox{$p$+A}}

\newcommand{\nch}{\lr{N_{\mathrm{ch}}}}

\newcommand{\nchb}{\mbox{$N_{\mathrm{ch}}^{\mathrm{sel1}}$}}
\newcommand{\nchc}{\mbox{$N_{\mathrm{ch}}^{\mathrm{sel2}}$}}
\newcommand{\nchrec}{\mbox{$N_{\mathrm{ch}}^{\mathrm{rec}}$}}

\newcommand{\sqrtsnn}{\mbox{$\sqrt{s_{\mathrm{NN}}}$}}
\newcommand{\Dphi}{\mbox{$\Delta \phi$}}
\newcommand{\Deta}{\mbox{$\Delta \eta$}}

\newcommand{\scn}{\mathrm{sc}_{n,m}\{4\}}
\newcommand{\sca}{\mathrm{sc}_{2,3}\{4\}}
\newcommand{\scb}{\mathrm{sc}_{2,4}\{4\}}

\newcommand{\nsca}{\mathrm{nsc}_{2,3}\{4\}}
\newcommand{\nscb}{\mathrm{nsc}_{2,4}\{4\}}
\newcommand{\acn}{\mathrm{ac}_n\{3\}}
\newcommand{\aca}{\mathrm{ac}_2\{3\}}

\newcommand{\naca}{\mathrm{nac}_2\{3\}}

\hypersetup{pdftitle={ATLAS document},pdfauthor={The ATLAS Collaboration}}
\begin{document}

\maketitle

\section{Introduction}
\label{intro}
One of the goals in the studies of azimuthal correlations in high-energy nuclear collisions at the Relativistic Heavy Ion Collider (RHIC) and the Large Hadron Collider (LHC) is to understand the multi-parton dynamics of QCD in the strongly coupled non-perturbative regime~\cite{Shuryak:2014zxa}. Measurements of azimuthal correlations in small collision systems, such as $pp$, $\pA$ or $d$+A collisions, have revealed the ridge phenomenon~\cite{CMS:2012qk,Abelev:2012ola,Aad:2012gla,Aad:2014lta,Khachatryan:2015waa}: enhanced production of particle pairs at small azimuthal angle separation, $\Dphi$, extended over a wide range of pseudorapidity separation, $\Deta$. The azimuthal structure has been related to harmonic modulation of particle densities, characterised by a Fourier expansion, d$N/{\textrm d}\phi\propto 1+2\sum_{n=1}^{\infty}v_{n}\cos n(\phi-\Phi_{n})$, where $v_n$ and $\Phi_n$ represent the magnitude and the event-plane angle of the $n^{\mathrm{th}}$-order flow harmonic. They are also conveniently represented by the flow vector: $ {\bm V}_n = v_n e^{{\textrm i}n\Phi_n}$. The $v_n$ are known to depend on the collision system, but have weak dependence on collision energies~\cite{Aaboud:2016yar,Khachatryan:2015waa}. The ridge reflects multi-parton dynamics early in the collision and has generated significant interest in the high-energy physics community. A key question is whether the long-range multi-particle collectivity reflects initial momentum correlation from gluon saturation effects~\cite{Dusling:2013qoz}, or a final-state hydrodynamic response to the initial transverse collision geometry~\cite{Bozek:2013uha}. 

Further insight into the ridge phenomenon is obtained via a multi-particle correlation technique, known as cumulants, involving three or more particles~\cite{Borghini:2000sa,Bilandzic:2010jr,Bilandzic:2013kga}. The multi-particle cumulants probe the event-by-event fluctuation of a single flow harmonic $v_n$, as well as the correlated fluctuations between two flow harmonics, $v_n$ and $v_m$. These event-by-event fluctuations are often represented by probability density distributions $p(v_n)$ and $p(v_n, v_m)$, respectively. For instance, the four-particle cumulants $c_n\{4\}=\lr{v_n^4}-2\lr{v_n^2}^2$ constrain the width of $p(v_n)$~\cite{Borghini:2000sa}, while the four-particle symmetric cumulants $\scn=\lr{v_n^2v_m^2}-\lr{v_n^2}\lr{v_m^2}$ quantify the lowest-order correlation between $v_n$ and $v_m$~\cite{Bilandzic:2013kga}. The three-particle asymmetric cumulants such as $\acn=\lr{{\bm V}_n^2{\bm V}_{2n}^*}=\lr{v_n^2v_{2n}\cos2n(\Phi_n-\Phi_{2n})}$~\cite{Aad:2014lta,Jia:2017hbm} are sensitive to correlations involving both the flow magnitude $v_n$ and flow phase $\Phi_n$.

One of the challenges in the study of azimuthal correlations in small collision systems is how to distinguish the long-range ridge from ``non-flow'' correlations involving only a few particles, such as resonance decays, jets, or dijets. For two-particle correlations, the non-flow contribution is commonly suppressed by requiring a large $\Deta$ gap between the two particles in each pair and a peripheral subtraction procedure~\cite{Abelev:2012ola,Aad:2012gla,Aad:2014lta,Aad:2015gqa,Aaboud:2016yar,Khachatryan:2016txc}. For multi-particle cumulants, the non-flow contributions can be suppressed by requiring correlation between particles from different subevents separated in $\eta$, while preserving the genuine long-range multi-particle correlations associated with the ridge. Here each subevent is a collection of particles in a given $\eta$ range. This so-called ``subevent method'' has been demonstrated to measure reliably $c_n\{4\}$ and $\scn$~\cite{Jia:2017hbm,Huo:2017nms}. In contrast, $c_n\{4\}$ and $\scn$ based on the standard cumulant method are contaminated by non-flow correlations over the full multiplicity range in $pp$ collisions and the low multiplicity region in $\pA$ collisions~\cite{Huo:2017nms}. In small collision systems, measurements have been performed for $c_n\{4\}$ with both the standard~\cite{Khachatryan:2016txc,Aaboud:2017acw} and subevent methods~\cite{Aaboud:2017blb}, and for $\scn$ with the standard method~\cite{Sirunyan:2017uyl}. The subevent method has not yet been used to measure $\scn$, and no measurements of $\acn$ have ever been attempted in small collision systems.

This Letter presents measurements of $\sca$,  $\scb$ and $\aca$ in $pp$ collisions at $\sqrt{s}=$ 13~TeV, $p$+Pb collisions at $\sqrtsnn=$ 5.02~TeV and low-multiplicity Pb+Pb collisions at $\sqrtsnn=$ 2.76~TeV. They are obtained using two-, three- and four-subevent cumulant methods and are compared with results from the standard cumulant method. The cumulants are normalised by the $\lr{v_n^2}$ obtained from a two-particle correlation analysis~\cite{Aaboud:2016yar} to quantify their relative correlation strength. The measurements suggest that the results obtained with the standard method are strongly contaminated by correlations from non-flow sources. The results obtained with the three-subevent method or the four-subevent method provide new evidence of long-range three- or four-particle azimuthal correlations.

The Letter is organised as follows. Details of the ATLAS detector, the trigger system, datasets, as well as event and track selections are provided in Sections~\ref{sec:data} to~\ref{sec:sel}. Section~\ref{sec:method} describes the standard and subevent cumulant methods used in this analysis. The analysis procedure and systematic uncertainties are described in Sections~\ref{sec:ana} and~\ref{sec:syscheck}, respectively. The measured cumulants are presented in Section~\ref{sec:result}. A summary is given in Section~\ref{sec:sum}.

\section{Detector and trigger}
\label{sec:data}
The ATLAS detector~\cite{Aad:2008zzm} provides nearly full solid-angle coverage around the collision point with tracking detectors, calorimeters, and muon chambers, and is well suited for measurement of multi-particle correlations over a large pseudorapidity range.\footnote{ATLAS typically uses a right-handed coordinate system with its origin at the nominal interaction point (IP) in the centre of the detector and the $z$-axis along the beam pipe. The $x$-axis points from the IP to the centre of the LHC ring, and the $y$-axis points upward. Cylindrical coordinates $(r,\phi)$ are used in the transverse plane, $\phi$ being the azimuthal angle around the beam pipe. By default, the pseudorapidity is defined in terms of the polar angle $\theta$ as $\eta=-\ln\tan(\theta/2)$. However, for asymmetric $\pPb$ or Pb+$p$ collisions, the $-z$ direction is always defined as the direction of the Pb beam.} The measurements were performed using primarily the inner detector (ID), minimum-bias trigger scintillators (MBTS) and the zero-degree calorimeters (ZDC). The ID detects charged particles within $|\eta| < 2.5$ using a combination of a silicon pixel detector, a silicon microstrip detector~(SCT), and a straw-tube transition radiation tracker, all immersed in a 2 T axial magnetic field~\cite{Aad:ID}. An additional pixel layer, the ``insertable B-layer'' (IBL)~\cite{ATLAS-TDR-19} is installed between the Run-1 (2010--2013) and Run-2 (2015--2018) periods. The MBTS detects charged particles within $2.1\lesssim|\eta|\lesssim3.9$ using two hodoscopes of counters positioned at $z = \pm$~3.6~m. The ZDC, used only in $\pPb$ and Pb+Pb collisions, are positioned at $\pm$140~m from the collision point, and detect neutral particles, primarily neutrons and photons, with $|\eta|>8.3$. 

The ATLAS trigger system~\cite{Aad:2012xs,Aaboud:2016leb} consists of a first-level (L1) trigger implemented using a combination of dedicated electronics and programmable logic, and a high-level trigger (HLT) implemented in processors.  The HLT reconstructs charged-particle tracks using methods similar to those applied in the offline analysis. The HLT enables the high-multiplicity track triggers (HMT) to select events according to the number of tracks having $\pT > 0.4$~GeV matched to the primary vertex, $N_{\mathrm{ch}}^{\mathrm{HLT}}$. The different HMT triggers apply additional requirements on either the total transverse energy ($\et$) in the calorimeters or the number of hits in the MBTS found by the L1 trigger, as well as on $N_{\mathrm{ch}}^{\mathrm{HLT}}$ by the HLT trigger. The $pp$ and $\pPb$ data were collected using combinations of the minimum-bias and HMT triggers. The minimum-bias trigger required either a hit in at least one MBTS counter, or a hit in at least one MBTS counter on each side, or at least one reconstructed track at the HLT seeded by a random trigger at L1. More detailed information about the triggers used for the $pp$ and $\pPb$ data and their performance can be found in Refs.~\cite{Aad:2016mok,Aaboud:2016yar} and Refs.~\cite{Aad:2014lta,atlas:3}, respectively.

\section{Datasets and Monte Carlo simulations}
\label{sec:datamc}
This analysis is based on ATLAS datasets corresponding to integrated luminosities of 0.9 $\mathrm{pb}^{-1}$ of $pp$ data recorded at $\sqrt{s} = 13$~TeV, 28~$\mathrm{nb}^{-1}$ of $\pPb$ data recorded at $\sqrtsnn = 5.02$~TeV, and 7~$\mu$b$^{-1}$ of Pb+Pb data at $\sqrtsnn = 2.76$~TeV. The $2.76$~TeV Pb+Pb data were collected in 2010. The $\pPb$ data were mainly collected in 2013, but also include 0.3~$\mathrm{nb}^{-1}$ of data collected in 2016, which increase the number of events at moderate multiplicity (see Section~\ref{sec:sel}). During both $\pPb$ runs, the LHC was configured to provide a 4~TeV proton beam and a 1.57~TeV per-nucleon Pb beam, which produced collisions at $\sqrtsnn = 5.02$~TeV, with a rapidity shift of $0.465$ of the nucleon--nucleon centre-of-mass frame towards the proton beam direction relative to the ATLAS rest frame. The direction of the Pb beam is always defined to have negative pseudorapidity. The 13~TeV $pp$ data were collected during several special runs of the LHC with low pile-up in 2015 and 2016. A summary of the datasets used in this analysis is shown in Table~\ref{tab:1}.
\begin{table}[!h]
\centering
\caption{\label{tab:1} The list of datasets used in this analysis.}
\begin{tabular}{c|c|c|c}\hline 
                         &Pb+Pb&$\pPb$&$pp$\\\hline
\multirow{2}{*}{Integrated luminosity (year)}&7 $\mu$b$^{-1}$ (2010)&  28 nb$^{-1}$ (2013)  & 0.07 pb$^{-1}$ (2015)  \\\cline{2-4}
                                             &                      &  0.3 nb$^{-1}$ (2016) & 0.84 pb$^{-1}$ (2016) \\\hline
\end{tabular}
\end{table}

The track reconstruction efficiency was determined using simulated Monte Carlo (MC) event samples~(Section~\ref{sec:sel}). The $pp$ events were simulated with the $\PYTHIA{8}$ MC event generator~\cite{Sjostrand:2007gs} using the A2 set of tuned parameters with MSTW2008LO parton distribution functions~\cite{atlas:6}. The HIJING event generator~\cite{Gyulassy:1994ew} was used to produce Pb+Pb and $\pPb$ collisions with the same energy and the same boost of the centre-of-mass system as in the data. The detector response was simulated using $\GEANT{4}$~\cite{Agostinelli:2002hh,Aad:2010ah} with detector conditions matching those during the data-taking. The simulated events and data events are reconstructed with the same algorithms. The MC sample for Pb+Pb events in the multiplicity region of interest is very small, and so the track reconstruction efficiency for Pb+Pb was taken from the larger $\pPb$ sample reconstructed with the same reconstruction algorithm. The efficiency in $\pPb$ events was found to be consistent with the efficiency from the Pb+Pb MC simulation~\cite{Aaboud:2017acw}. 

\section{Event and track selection}
\label{sec:sel}
The offline event selection for the $pp$ and $\pPb$ data requires at least one reconstructed vertex with its longitudinal position satisfying $|\zvtx|< 100$~mm relative to the nominal interaction point. The vertex is required to have at least two associated tracks with $\pT>0.4$~GeV. The mean number of collisions per bunch crossing, $\mu$, was 0.002--0.8 for the 13 TeV $pp$ data, 0.03 for the 2013 $\pPb$ data, and 0.001--0.006 for the 2016 $\pPb$ data. In order to suppress additional interactions in the same bunch crossing (referred to as pile-up) in $pp$ collisions, events containing additional vertices with at least four associated tracks are rejected. In $p$+Pb collisions, events with more than one good vertex, defined as any vertex for which the scalar sum of the $\pT$ of the associated tracks is greater than 5 GeV, are rejected. The remaining pile-up events are further suppressed by using the signal in the ZDC in the direction of the Pb beam. This signal is calibrated to the number of detected neutrons, $N_{n}$, by using the location of the peak corresponding to a single neutron. The distribution of $N_{n}$ in events with pile-up is broader than that for the events without pile-up. Hence a simple requirement on the ZDC signal distribution is used to further suppress events with pile-up, while retaining more than 98\% of events without pile-up. The impact of residual pile-up, at the level of $\lesssim 10^{-3}$, is studied by comparing the results obtained from data with different $\mu$ values.

The offline event selection for the Pb+Pb data requires $|\zvtx|< 100$~mm. The selection also requires a time difference $|\Delta t| < 3$ ns between signals in the MBTS trigger counters on either side of the interaction point to suppress non-collision backgrounds. A coincidence between the ZDC signals at forward and backward pseudorapidity is required to reject a variety of background processes, while maintaining high efficiency for inelastic processes. The fraction of events with more than one interaction after applying these selection criteria is less than $10^{-4}$.

Charged-particle tracks and collision vertices are reconstructed using algorithms optimised for improved performance for Run-2. In order to compare directly with the $pp$ and $\pPb$ systems using event selections based on the multiplicity of the collisions, a subset of data from low-multiplicity Pb+Pb collisions, collected during the 2010 LHC heavy-ion run with a minimum-bias trigger, was analysed using the same track reconstruction algorithm as that used for $\pPb$ collisions. For the Pb+Pb and 2013 $p$+Pb analyses, tracks are required to have a $\pT$-dependent minimum number of hits in the SCT. The transverse~($d_0$) and longitudinal ($z_0$ $\sin\theta$) impact parameters of the track relative to the vertex are required to be less than 1.5~mm.  Additional requirements $|d_0|/\sigma_{d_0} < 3$ and $|z_0\sin\theta|/\sigma_{z_0}< 3$ are imposed, where $\sigma_{d_0}$ and $\sigma_{z_0}$ are the uncertainties of the transverse and longitudinal impact parameter values, respectively. A more detailed description of the track selection for the 2010 Pb+Pb data and 2013 $\pPb$ data can be found in Refs.~\cite{Aad:2014lta,Aaboud:2017acw}.

For all the data taken since the start of Run-2, the track selection criteria make use of the IBL, as described in Refs.~\cite{Aad:2016mok,Aad:2015gqa}. For the $pp$ and 2016 $p$+Pb analyses, the tracks are required to satisfy $|d_0^{\mathrm{BL}}|<1.5$ mm and $|z_0\sin\theta|<1.5$~mm, where $d_0^{\mathrm{BL}}$ is the transverse impact parameter of the track relative to the beam line~(BL).

The cumulants are calculated using tracks passing the above selection requirements, and having $|\eta|<2.5$ and $0.3<\pT <3$~GeV or $0.5<\pT <5$~GeV. These two $\pT$ ranges are chosen because they were often used in the previous ridge measurements at the LHC~\cite{Khachatryan:2015waa,Khachatryan:2016txc,Aad:2015gqa,Aaboud:2016yar,Aaboud:2017acw}. However, to count the number of reconstructed charged particles for event-class definition (denoted by $\nchrec$), tracks with $\pT>0.4$~GeV and $|\eta|<2.5$ are used for compatibility with the requirements in the HLT selections described above. Due to different trigger requirements, most of the $\pPb$ events with $\nchrec>150$ are provided by the 2013 dataset, while the 2016 dataset provides most of the events at lower $\nchrec$.

The efficiency of the combined track reconstruction and track selection requirements is estimated using MC samples reconstructed with the same algorithms and selection requirements as in data. Efficiencies, $\epsilon(\eta,\pT)$, are evaluated as a function of track $\eta$, $\pT$ and the number of reconstructed charged-particle tracks, but averaged over the full range in azimuth.  The efficiencies are similar for events with the same multiplicity. For all collision systems, the efficiency increases by about 4\% as track $\pT$ increases from 0.3~GeV to 0.6 GeV. Above 0.6 GeV, the efficiency is independent of $\pT$ and reaches 86\% (72\%) for Run-1 $pp$ and $\pPb$, and 83\% (70\%) for Pb+Pb and Run-2 $\pPb$ collisions, at $\eta\approx 0$ ($|\eta| > 2$). The efficiency is independent of the event multiplicity for $\nchrec > 40$. For lower-multiplicity events the efficiency is smaller by up to 3\% due to broader $d_0$ and $z_0\sin\theta$ distributions~\cite{Aaboud:2017acw}.

The fraction of falsely reconstructed charged-particle tracks is also estimated and found to be negligibly small in all datasets. This fraction decreases with increasing track $\pT$, and even at the lowest transverse momenta of 0.3~GeV it is below 1\% of the total number of tracks. Therefore, there is no correction for the presence of such tracks in the analysis. 

In the simulated events, the reconstruction efficiency reduces the measured charged-particle multiplicity relative to the generated multiplicity for primary charged particles. A correction factor $b$ is used to correct $\nchrec$ to obtain the efficiency-corrected average number of charged particles per event, $\nch=b\lr{\nchrec}$. The value of the correction factor is obtained from the MC samples described above, and is found to be nearly independent of $\nchrec$ in the range used in this analysis, $\nchrec<400$. Its value and the associated uncertainties are $b=1.29$ $\pm$ 0.05 for the Pb+Pb and 2013 $\pPb$ collisions and $b=1.18\pm$~0.05 for Run-2 $\pPb$ and $pp$ collisions~\cite{Aaboud:2016jnr}. Both $\scn$ and $\aca$ are then studied as a function of $\nch$.

\section{Cumulant method}
\label{sec:method}
The multi-particle cumulant method~\cite{Borghini:2000sa} has the advantage of directly reducing non-flow correlations from jets and dijets. The mathematical framework for the standard cumulant is based on the Q-cumulants discussed in Refs.~\cite{Bilandzic:2010jr,Bilandzic:2013kga,DiFrancesco:2016srj}. It was extended recently to the case of subevent cumulants in Refs.~\cite{Jia:2017hbm,Huo:2017nms}. These methods are briefly summarised below.

\subsection{Cumulants in the standard method}
The standard cumulant method calculates $k$-particle azimuthal correlations, $\lr{\{k\}}$, in one event using a complex number notation~\cite{Bilandzic:2010jr,Bilandzic:2013kga}:
\begin{eqnarray}\label{eq:1}
\lr{\{2\}_n} = \lr{{\mathrm{e}}^{{\mathrm i}n(\phi_1-\phi_2)}}\;,\;\;\lr{\{3\}_{n}}= \lr{{\mathrm{e}}^{{\mathrm i}n(\phi_1+\phi_2-2\phi_3)}}\;,\;\;\lr{\{4\}_{n,m}}= \lr{{\mathrm{e}}^{{\mathrm i}n(\phi_1-\phi_2)+{\mathrm i}m(\phi_3-\phi_4)}}\;,
\end{eqnarray}
where ``$\lr{}$'' denotes a single-event average over all pairs, triplets or quadruplets, respectively. The averages from Eq.~\eqref{eq:1} can be expressed in terms of per-particle normalised flow vectors ${\bm q}_{n;l}$ with $l=1,2...$ in each event~\cite{Bilandzic:2010jr}: 
\begin{eqnarray}\label{eq:2}
{\bm q}_{n;l} \equiv {\sum\limits_j  w_j^l{\mathrm{e}}^{{\textrm i}n\phi_j}}{\left.\vphantom{\sum\limits_j}\right/}{\sum\limits_jw_j^l}\;, 
\end{eqnarray}
where the sum runs over all tracks in the event and $w_j$ is a weight assigned to the $j^{\textrm{th}}$ track. This weight is constructed to correct for both detector non-uniformity and tracking inefficiency as explained in Section~\ref{sec:ana}. 

The multi-particle asymmetric and symmetric cumulants are obtained from $\lr{\{k\}}$ as:
\begin{eqnarray}\label{eq:3}
\acn = \llrr{\{3\}_{n}}\;,\;\;\scn = \llrr{\{4\}_{n,m}}-\llrr{\{2\}_{n}}\llrr{\{2\}_{m}}\;,
\end{eqnarray}
where ``$\llrr{}$'' represents a weighted average of $\lr{\{k\}}$ over an event ensemble with similar $\nchrec$. One averages first over all distinct pairs, triplets or quadruplets in one event to obtain $\lr{\{2\}_{n}}$, $\lr{\{2\}_{m}}$, $\lr{\{3\}_{n}}$ and $\lr{\{4\}_{n,m}}$. Then the obtained values are averaged over an event ensemble with similar $\nchrec$ to obtain $\scn$ and $\acn$. In the absence of non-flow correlations, $\scn$ and $\acn$ measure the correlation between $v_n$ and $v_m$ or between $v_n$ and $v_{2n}$:
\begin{eqnarray}\label{eq:3b}
\acn&=&\lr{{\bm V}_n^2{\bm V}_{2n}^*}=\lr{v_n^2v_{2n}\cos2n(\Phi_n-\Phi_{2n})}\;,\\
\scn&=&\lr{v_n^2v_m^2}-\lr{v_n^2}\lr{v_m^2}\;,
\end{eqnarray}
where the averages are taken over the events. This analysis measures three types of cumulants defined in Eq.~\eqref{eq:3}: $\sca$, $\scb$ and $\aca$.

\subsection{Cumulants in the subevent method}
In the standard cumulant method described above, all $k$-particle multiplets involved in $\lr{\{k\}_n}$ and $\lr{\{k\}_{n,m}}$ are selected using tracks in the entire ID acceptance of $|\eta|<\mathrm{\eta_{max}}=2.5$. To suppress further the non-flow correlations that typically involve a few particles within a localised region in $\eta$, the tracks are divided into several subevents, each covering a unique $\eta$ interval. The multi-particle correlations are then constructed by only correlating tracks between different subevents. 

In the two-subevent cumulant method, the tracks are divided into two subevents, labelled by $a$ and $b$, according to $-\mathrm{\eta_{max}}<\eta_a<0$ and $0\leq\eta_b<\mathrm{\eta_{max}}$. The per-event $k$-particle azimuthal correlations are evaluated as:
\begin{eqnarray}\nonumber
\lr{\{2\}_n}_{a|b} = \lr{{\mathrm{e}}^{{\mathrm i}n(\phi_1^a-\phi_2^b)}},\;\lr{\{3\}_{n}}_{2a|b}= \lr{{\mathrm{e}}^{{\mathrm i}n(\phi_1^a+\phi_2^a-2\phi_3^b)}},\;\lr{\{4\}_{n,m}}_{2a|2b}= \lr{{\mathrm{e}}^{{\mathrm i}n(\phi_1^a-\phi_2^b)+{\mathrm i}m(\phi_3^a-\phi_4^b)}},\\\label{eq:4}
\end{eqnarray}
where the superscript or subscript $a$ ($b$) indicates tracks chosen from the subevent $a$ ($b$). Here the three- and four-particle cumulants are defined as:
\begin{eqnarray}\nonumber
\mathrm{ac}_n^{2a|b}\{3\} = \llrr{\{3\}_{n}}_{2a|b}\;,\;\mathrm{sc}_{n,m}^{2a|2b}\{4\}  = \llrr{\{4\}_{n,m}}_{2a|2b}-\llrr{\{2\}_{n}}_{a|b}\llrr{\{2\}_{m}}_{a|b}\;.
\end{eqnarray}
The two-subevent method suppresses correlations within a single jet (intra-jet correlations), since particles from one jet usually fall in one subevent.

In the three-subevent cumulant method, tracks in each event are divided into three subevents $a$, $b$ and $c$, each covering one third of the $\eta$ range,  $-\mathrm{\eta_{max}}<\eta_a<-\mathrm{\eta_{max}}/3$, $|\eta_b|\leq\mathrm{\eta_{max}}/3$ and $\mathrm{\eta_{max}}/3<\eta_c<\mathrm{\eta_{max}}$. The multi-particle azimuthal correlations and cumulants are then evaluated as:
\begin{eqnarray}\label{eq:6a}
\lr{\{3\}_{n}}_{a,b|c}= \lr{{\mathrm{e}}^{{\mathrm i}n(\phi_1^a+\phi_2^b-2\phi_3^c)}}\;,\;\lr{\{4\}_{n,m}}_{a,b|2c}= \lr{{\mathrm{e}}^{{\mathrm i}n(\phi_1^a-\phi_2^c)+{\mathrm i}m(\phi_3^b-\phi_4^c)}}\;,
\end{eqnarray}
and
\begin{eqnarray}\label{eq:6c}
\mathrm{ac}_n^{a,b|c}\{3\} = \llrr{\{3\}_{n}}_{a,b|c}\;,\;\;\mathrm{sc}_{n,m}^{a,b|2c}\{4\}  = \llrr{\{4\}_{n,m}}_{a,b|2c}-\llrr{\{2\}_{n}}_{a|c}\llrr{\{2\}_{m}}_{b|c}\;.
\end{eqnarray}
Since a dijet event usually produces particles in at most two subevents, the three-subevent method efficiently suppresses the non-flow contribution from inter-jet correlations associated with dijets. To maximise the statistical precision, the $\eta$ range for subevent $a$ is swapped with that for subevent $b$ or $c$, and the results are averaged to obtain the final values.

The four-subevent cumulant method is only relevant for the symmetric cumulants $\scn$. Tracks in each event are divided into four subevents $a$, $b$, $c$, and $d$, each covering one quarter of the $\eta$ range: $-\mathrm{\eta_{max}}<\eta_a<-\mathrm{\eta_{max}}/2$, $-\mathrm{\eta_{max}}/2\leq\eta_b<0$, $0\leq\eta_c<\mathrm{\eta_{max}}/2$, and $\mathrm{\eta_{max}}/2\leq\eta_d<\mathrm{\eta_{max}}$. The multi-particle azimuthal correlations and cumulants are then evaluated as:
\begin{eqnarray}\label{eq:7a}
&&\lr{\{4\}_{n,m}}_{a,b|c,d}= \lr{{\mathrm{e}}^{{\mathrm i}n(\phi_1^a-\phi_2^c)+{\mathrm i}m(\phi_3^b-\phi_4^d)}}\;,\\\label{eq:7b}
&&\mathrm{sc}_{n,m}^{a,b|c,d}\{4\}  = \llrr{\{4\}_{n,m}}_{a,b|c,d}-\llrr{\{2\}_{n}}_{a|c}\llrr{\{2\}_{m}}_{b|d}\;.
\end{eqnarray}
The four-subevent method based on Eqs.~\eqref{eq:7a} and \eqref{eq:7b} should further suppress the residual non-flow contributions, for instance when each of the two jets from the dijet falls across the boundary between two neighbouring subevents. To maximise the statistical precision, the $\eta$ ranges for the four subevents are swapped with each other, and the results are averaged to obtain the final values.

\subsection{Normalised cumulants}

Although the cumulants reflect the nature of the correlation between $v_n$ and $v_m$, their magnitudes also depend on the square of single flow harmonics $v_n^2$ and $v_m^2$, see Eq.~\eqref{eq:3b}. The dependence on the single flow harmonics can be scaled out via the normalised cumulants~\cite{Giacalone:2016afq,Das:2017ned}:
\begin{eqnarray}\label{eq:8a}
\nsca&=&\frac{\sca}{v_2\{2\}^2v_3\{2\}^2}= \frac{\lr{v_2^2v_3^2}}{\lr{v_2^2}\lr{v_3^2}}-1\;,\\\label{eq:8b}
\nscb&=&\frac{\scb}{v_2\{2\}^2v_4\{2\}^2}= \frac{\lr{v_2^2v_4^2}}{\lr{v_2^2}\lr{v_4^2}}-1\;,\\\label{eq:8c}
\mathrm{nac}_2\{3\}&=&\frac{\aca}{\sqrt{\left(2v_2\{2\}^4+c_2\{4\}\right)c_{4}\{2\}}} = \frac{\lr{v_2^2v_{4}\cos4(\Phi_2-\Phi_{4})}}{\sqrt{\lr{v_2^4}\lr{v_{4}^2}}}\;,
\end{eqnarray}
where the $v_n\{2\}^2=\lr{v_n^2}$ are flow harmonics obtained using a two-particle correlation method based on a peripheral subtraction technique~\cite{Aad:2015gqa,Aaboud:2016yar}, and $c_2\{4\}=\lr{v_2^4}-2\lr{v_2^2}^2$ are four-particle cumulant results from Refs.~\cite{Aaboud:2017blb,Aaboud:2017acw}. This definition for $\mathrm{nac}_2\{3\}$ is motivated by Ref.~\cite{Bhalerao:2013ina}.

\section{Analysis procedure}
\label{sec:ana}
The measurement of the $\scn$ and $\aca$ follows the same analysis procedure as for the four-particle cumulants $c_n\{4\}$ in Ref.~\cite{Aaboud:2017blb}. The multi-particle cumulants are calculated in three steps using charged particles with $|\eta|<2.5$. In the first step, $\lr{\{2\}_n}$, $\lr{\{3\}_n}$ and $\lr{\{4\}_{n,m}}$ from Eqs.~\eqref{eq:1},~\eqref{eq:4}, \eqref{eq:6a} and \eqref{eq:7a} are calculated for each event from particles in one of two different $\pT$ ranges, $0.3<\pT<3$ GeV and $0.5<\pT<5$ GeV. The numbers of reconstructed charged particles in these $\pT$ ranges are denoted by $\nchb$ and $\nchc$, respectively. 

In the second step, the correlators $\lr{\{k\}}$ for $0.3<\pT<3$ GeV ($0.5<\pT<5$ GeV) are averaged over events with the same $\nchb$ ($\nchc$) to obtain $\llrr{\{k\}}$, and then $\sca$, $\scb$ and $\aca$. The $\sca$, $\scb$ and $\aca$ values are then averaged in broader multiplicity ranges of the event ensemble, weighted by number of events, to obtain statistically significant results.

In the third step, the $\sca$, $\scb$ and $\aca$ values obtained for a given $\nchb$ or $\nchc$ are mapped to $\lr{\nchrec}$, the average number of reconstructed charged particles with $\pT>0.4$ GeV. The mapping procedure is necessary so that $\sca$, $\scb$ and $\aca$ obtained for the two different $\pT$ ranges can be compared using a common $x$-axis defined by $\lr{\nchrec}$. The $\lr{\nchrec}$ value is then converted to $\nch$, the efficiency-corrected average number of charged particles with $\pT > 0.4$ GeV, as discussed in Section~\ref{sec:sel}.

In order to account for detector inefficiencies and non-uniformity, particle weights used in Eq.~\eqref{eq:2} are defined as:
\begin{eqnarray}\nonumber
w(\phi,\eta,\pT) = d(\phi,\eta)/\epsilon(\eta,\pT)\;.
\end{eqnarray}
The additional weight factor $d(\phi,\eta)$ accounts for non-uniformities in the azimuthal acceptance of the detector as a function of $\eta$. All reconstructed charged particles with $\pT>0.3$ GeV are entered into a two-dimensional histogram $N(\phi,\eta)$, and the weight factor is then obtained as $d(\phi,\eta) \equiv \lr{N(\eta)}/N(\phi,\eta)$, where $\lr{N(\eta)}$ is the track density averaged over $\phi$ in the given $\eta$ bin. This procedure removes most of the $\phi$-dependent non-uniformity in the detector acceptance~\cite{Aaboud:2017acw}.

In order to calculate the normalised cumulants from Eqs.~\eqref{eq:8a}--\eqref{eq:8c}, the flow harmonics $v_n\{2\}$ are obtained from a ``template fit'' of two-particle $\Delta\phi$ correlation as described in Refs.~\cite{Aad:2015gqa,Aaboud:2016yar}. The $v_n\{2\}$ values are calculated identically to the procedure used in the previous ATLAS publications~\cite{Aad:2015gqa,Aaboud:2016yar}, but are further corrected for a bias, which exists only if $v_n\{2\}$ changes with $\nchrec$. The details of the correction procedure are given in the Appendix~\ref{sec:app1} and are discussed briefly below.

The standard procedure of Refs.~\cite{Aad:2015gqa,Aaboud:2016yar} first constructs a $\Delta\phi$ distribution for pairs of tracks with $|\Deta|>2$: the per-trigger-particle yield $Y(\Dphi)$ for a given $\nchrec$ range. The dominating non-flow jet peak at $\Dphi\sim\pi$ is estimated using low-multiplicity events with $\nchrec<20$ and separated via a template fit procedure, and the harmonic modulation of the remaining component is taken as the $v_n\{2\}^2$~\cite{Aaboud:2016yar}:
\begin{eqnarray}\nonumber
Y(\Dphi) = F\;Y(\Dphi)^{\mathrm{peri}} + G^{\mathrm{tmp}} \left(1+2\sum_{n=2}^{\infty} v_n\{2,\mathrm{tmp}\}^2 \cos n\Dphi\right)\;,
\end{eqnarray}
where superscripts ``peri'' and ``tmp'' indicate quantities for the $\nchrec<20$ event class and quantities after the template fit for the event class of interest, respectively. The scale factor $F$ and pedestal $G^{\mathrm{tmp}}$ are fixed by the fit, and $v_n\{2,\mathrm{tmp}\}$ are calculated from a Fourier transform. This procedure implicitly assumes that $v_n\{2\}$ is independent of $\nchrec$, and requires a small correction if $v_n\{2\}$ does change with $\nchrec$ (Appendix~\ref{sec:app1}). In $\pPb$ and Pb+Pb collisions, this correction in the $\nchrec>100$ region amounts to a 2--6\% reduction for $v_2\{2,\mathrm{tmp}\}$ and a 4--9\% reduction for $v_3\{2,\mathrm{tmp}\}$ and $v_4\{2,\mathrm{tmp}\}$. The correction is smaller for $v_2\{2,\mathrm{tmp}\}$ in $pp$ collisions as it is nearly independent of $\nchrec$~\cite{Aaboud:2016yar}.

\section{Systematic uncertainties}
\label{sec:syscheck}
The evaluation of the systematic uncertainties follows closely the procedure established for the four-particle cumulants $c_n\{4\}$ and described in Ref.~\cite{Aaboud:2017blb}. The main sources of systematic uncertainties are related to the detector azimuthal non-uniformity, track selection, track reconstruction efficiency, trigger efficiency and pile-up. Due to the relatively poor statistics and larger non-flow effects, the systematic uncertainties are typically larger in $pp$ collisions. The systematic uncertainties are also generally larger, in percentage, for four-particle cumulants $\scn$ than for the three-particle cumulants $\aca$, since the $|\scn|$ values are much smaller than those for $\aca$. The systematic uncertainties are generally similar among the two- and three- and four-subevent methods, but are different from those for the standard method, which is strongly influenced by non-flow correlations. The following discussion focuses on the three-subevent method, which is the default method used to present the final results.

The effect of detector azimuthal non-uniformity is accounted for using the weight factor $d(\phi,\eta)$. The impact of the weighting procedure is studied by fixing the weight to unity and repeating the analysis. The results are mostly consistent with the nominal results. The corresponding uncertainties for $\scn$ vary in the range of 0--4\%, 0--2\% and 1--2\% in $pp$, $\pPb$ and Pb+Pb collisions, respectively. The uncertainties for $\aca$ vary in the range of 0--2\% in $\pp$ collisions, and 0--1\% in $\pPb$ and Pb+Pb collisions, respectively. 

The systematic uncertainty associated with the track selection is estimated by tightening the $|d_0|$ and $|z_0\sin\theta|$ requirements. They are each varied from the default requirement of less than 1.5~mm to less than 1~mm. In $\pPb$ and Pb+Pb collisions, the requirement on the significance of impact parameters, $|d_0|/\sigma_{d_0}$ and $|z_0\sin\theta|/\sigma_{z_0}$ are also varied from less than 3 to less than 2. For each variation, the tracking efficiency is re-evaluated and the analysis is repeated. For $\aca$, which has a large flow signal, the differences from the nominal results are observed to be less than 2\% for all collision systems. For $\scn$, for which the signal is small, the differences from the nominal results are found to be in the range of 2--10\% in $pp$ collisions, 2--7\% in $\pPb$ collisions and 2--4\% in Pb+Pb collisions. The differences are smaller for results obtained for $0.5<\pT<5$ GeV than those obtained for $0.3<\pT<3$~GeV.

Previous measurements indicate that the azimuthal correlations (both the flow and non-flow components) have a strong dependence on $\pT$, but a relatively weak dependence on $\eta$~\cite{Aad:2014lta,Aaboud:2016yar}. Therefore, $\pT$-dependent systematic effects in the track reconstruction efficiency could affect the cumulant values. The uncertainty in the track reconstruction efficiency is mainly due to differences in the detector conditions and material description between the simulation and the data. The efficiency uncertainty varies between 1\% and 4\%, depending on track $\eta$ and $\pT$~\cite{Aaboud:2016yar,Aaboud:2017acw}. Its impact on multi-particle cumulants is evaluated by repeating the analysis with the tracking efficiency varied up and down by its corresponding uncertainty as a function of track $\pT$. For the standard cumulant method, which is more sensitive to jets and dijets, the evaluated uncertainty amounts to 2--6\% in $pp$ collisions and less than 2\% in $\pPb$ collisions for $\nch>100$. For the subevent methods, the evaluated uncertainty is typically less than 3\% for most of the $\nch$ ranges.

Most events in $pp$ and $\pPb$ collisions are collected with the HMT triggers with several online $\nchrec$ thresholds. In order to estimate the possible bias due to trigger inefficiency as a function of $\nch$, the offline $\nchrec$ requirements are changed such that the HMT trigger efficiency is at least 50\% or 80\%. The results are obtained independently for each variation. These results are found to be consistent with each other for the subevent methods, and show some differences for the standard cumulant method in the low $\nch$ region. The nominal analysis is performed using the 50\% efficiency selection and the differences between the nominal results and those from the 80\% efficiency selection are included in the systematic uncertainty. The changes for $pp$ collisions are in the range of 5--15\% for $\sca$, 2--8\% for $\scb$ and 1--5\% for $\aca$. The ranges for $\pPb$ collisions are much smaller due to the much sharper turn-on of the trigger efficiency and larger signal: they are estimated to be 1--3\% for $\sca$,  2--4\% for $\scb$ and 1--2\% for $\aca$.

In this analysis, a pile-up rejection criterion is applied to reject events containing additional vertices in $pp$ and $\pPb$ collisions. In order to check the impact of residual pile-up, the analysis is repeated without the pile-up rejection criterion. No differences are observed in $\pPb$ collisions, as is expected since the $\mu$ values in $\pPb$ are modest. For the 13~TeV $pp$ dataset, the differences with and without pile-up rejection are in the range of 0--7\% for $\sca$, 2--15\% for $\scb$ and 2--3\% for $\aca$. As a cross-check, the $pp$ data are divided into two samples with approximately equal number of events based on the $\mu$ value: $\mu>0.4$ and $\mu<0.4$, and the results are compared. No systematic differences are observed between the two independent datasets.

The systematic uncertainties from different sources are added in quadrature to determine the total systematic uncertainty. In $\pPb$ and Pb+Pb collisions, the total uncertainties are in the range of 3--8\% for $\sca$, 1--5\% for $\scb$ and 1--4\% for $\aca$. In $pp$ collisions, the total uncertainties are larger, mainly due to larger non-flow contribution, larger pile-up and the less sharp turn-on of the HMT triggers. They are in the ranges of 10--20\% for $\sca$, 10--20\% for $\scb$ and 2--5\% for $\aca$. The total systematic uncertainties are generally smaller than the statistical uncertainties.

The $v_n\{2\}$ values used to obtain normalised cumulants from Eqs.~\eqref{eq:8a}--\eqref{eq:8c} are measured following the prescription of the previous ATLAS publications~\cite{Aad:2015gqa,Aaboud:2016yar}, resulting in very similar systematic uncertainties. The correction for the bias of the template fit procedure, as described in Section~\ref{sec:ana}, reduces the sensitivity to the choice of the peripheral $\nchrec$ bin. The uncertainties of normalised cumulants are obtained by propagation of the uncertainties from the original cumulants and $v_n\{2\}$, taking into account that the correlated systematic uncertainties partially cancel out.

\section{Results}
\label{sec:result}
The results are presented in two parts. Section~\ref{sec:2} presents a detailed comparison between the standard method and subevent methods to demonstrate the ability of the subevent methods to suppress non-flow correlations. Section~\ref{sec:3} compares the cumulants among $pp$, $\pPb$ and Pb+Pb collisions to provide insight into the common nature of collectivity in these systems. 
\subsection{Comparison between standard and subevent methods}
\label{sec:2}
\begin{figure}[h!]
\begin{center}
\includegraphics[width=0.8\linewidth]{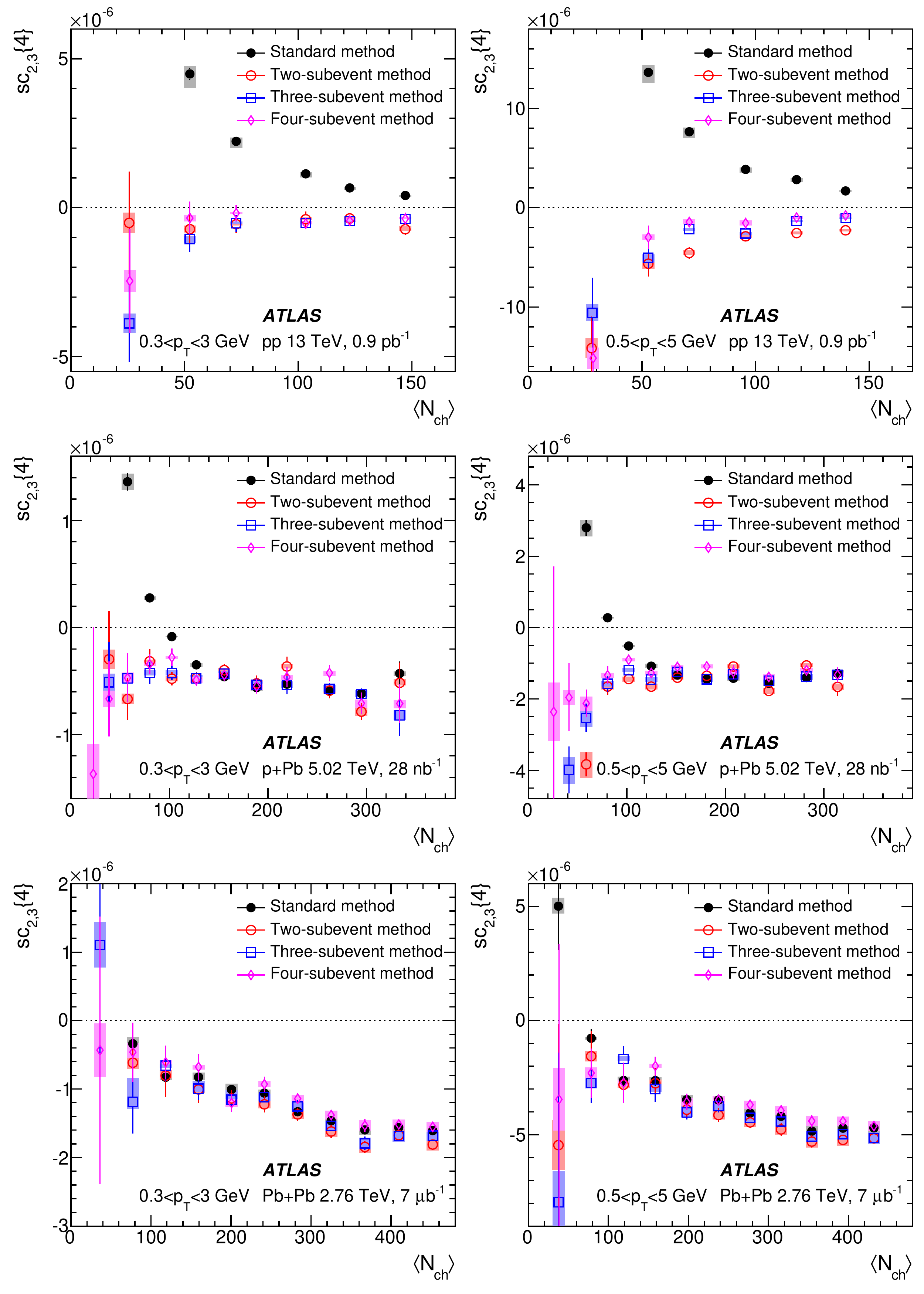}
\end{center}
\caption{\label{fig:b1} The symmetric cumulant $\sca$ as a function of $\nch$ for $0.3<\pT<3$~GeV (left panels) and $0.5<\pT<5$~GeV~(right panels) obtained for $pp$ collisions (top row), $\pPb$ collisions (middle row) and low-multiplicity Pb+Pb collisions (bottom row). In each panel, the $\sca$ is obtained from the standard method (filled symbol), the two-subevent method (open circles), three-subevent method (open squares) and four-subevent method (open diamonds). The error bars and shaded boxes represent the statistical and systematic uncertainties, respectively.}
\end{figure}

The top row of Figure~\ref{fig:b1} compares the $\sca$ values obtained from the standard, two-, three- and four-subevent methods from $pp$ collisions in $0.3<\pT<3$~GeV (left panel) and $0.5<\pT<5$~GeV (right panel). The values from the standard method are positive over the full $\nch$ range, and are larger at lower $\nch$ or in the higher $\pT$ range. This behaviour suggests that the $\sca$ values from the standard method in $pp$ collisions, including those from Ref.~\cite{Sirunyan:2017uyl}, are strongly influenced by non-flow effects in all $\nch$ and $\pT$ ranges~\cite{Huo:2017nms}. In contrast, the values from the subevent methods are negative over the full $\nch$ range, and they are slightly more negative at lowest $\nch$ and also more negative at higher $\pT$. The results are consistent among the various subevent methods for $0.3<\pT<3$~GeV. For the high $\pT$ region of $0.5<\pT<5$ GeV, results from the two-subevent method are systematically lower than those from the three- and four-subevent methods, suggesting that the two-subevent method may be affected by negative non-flow contributions. Such negative non-flow correlation has been observed in a $\PYTHIA{8}$ calculation~\cite{Huo:2017nms}.

The middle row of Figure~\ref{fig:b1} shows $\sca$ from $\pPb$ collisions. At $\nch>140$, the values are negative and consistent among all four methods, reflecting genuine long-range collective correlations. At $\nch<140$, the values are different between the standard method and the subevent methods. The $\sca$ from the standard method changes sign around $\nch\sim80$ and remains positive at lower $\nch$, reflecting the contribution from non-flow correlations. In contrast, the $\sca$ from various subevent methods are negative and consistent with each other at $\nch<140$, suggesting that they mainly reflect the genuine long-range correlations.

The bottom row of Figure~\ref{fig:b1} shows $\sca$ from Pb+Pb collisions. The results are consistent among all four methods across most of the $\nch$ range. In the low $\nch$ region, where the non-flow contribution is expected to be significant, the uncertainties of the results are too large to distinguish between different methods.

\begin{figure}[h!]
\begin{center}
\includegraphics[width=0.8\linewidth]{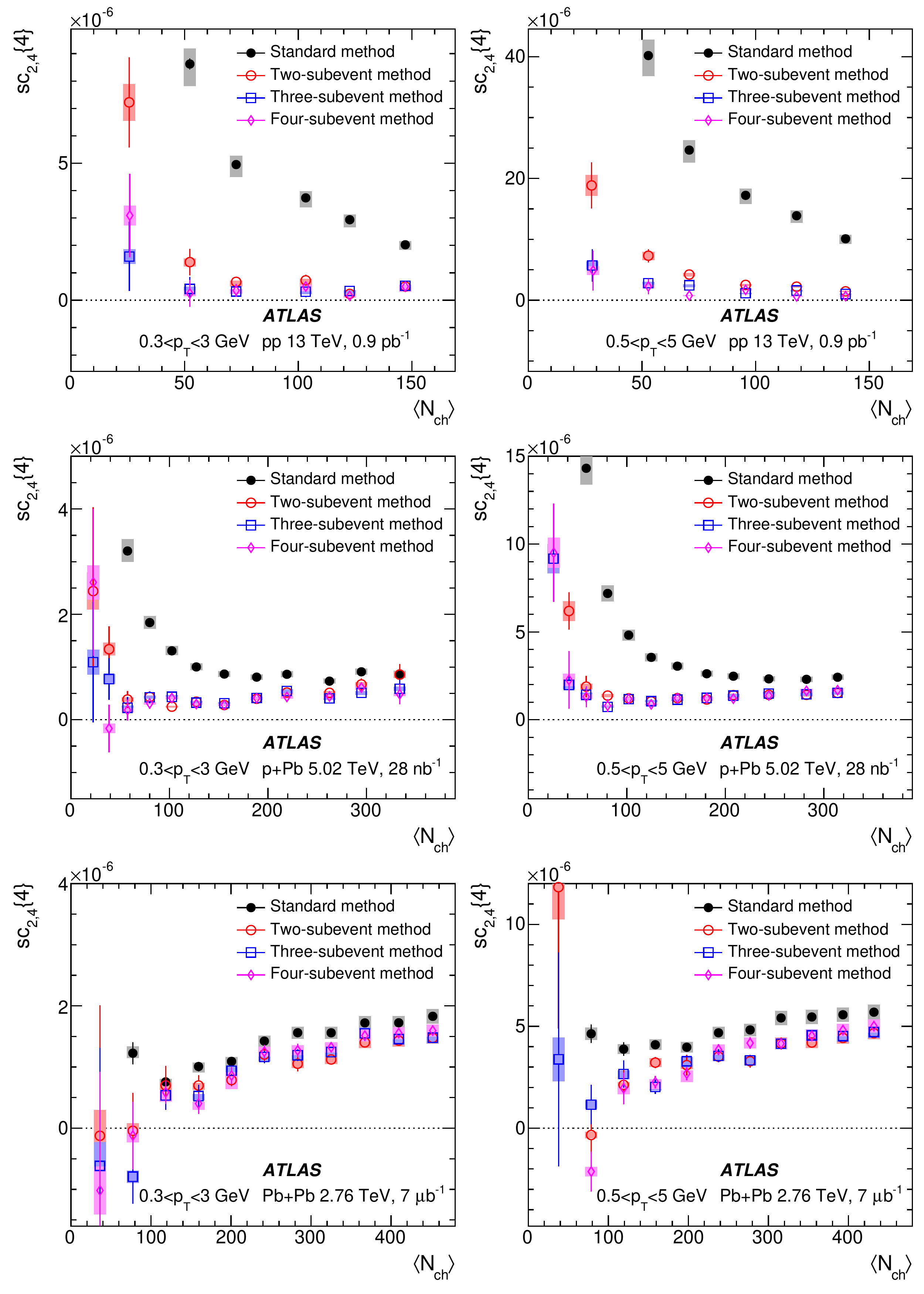}
\end{center}
\caption{\label{fig:b2} The symmetric cumulant $\scb$ as a function of $\nch$ for $0.3<\pT<3$ GeV (left panels) and $0.5<\pT<5$~GeV~(right panels) obtained for $pp$ collisions (top row), $\pPb$ collisions (middle row) and low-multiplicity Pb+Pb collisions (bottom row). In each panel, the $\scb$ is obtained from the standard method (filled symbol), two-subevent method (open circles), three-subevent method (open squares) and four-subevent method (open diamonds). The error bars and shaded boxes represent the statistical and systematic uncertainties, respectively.}
\end{figure}

The results for the symmetric cumulant $\scb$ are presented in Figure~\ref{fig:b2}. The top row shows the $\scb$ obtained from the standard, two-subevent, three-subevent and four-subevent methods from $pp$ collisions in $0.3<\pT<3$~GeV~(left panel) and $0.5<\pT<5$~GeV (right panel).  The values of $\scb$ are positive for all four methods. However, the results from the standard method are much larger than those from the subevent methods and also exhibit a much stronger increase towards the lower $\nch$ region. This behaviour is consistent with the expectation that the standard method is more affected by dijets. Significant differences are also observed between the two-subevent and three- or four-subevent methods at low $\nch$, but these differences decrease and disappear for $\nch>100$. Within the statistical uncertainties of the measurement, no differences are observed between the three- and four-subevent methods. This comparison suggests that the two-subevent method may not be sufficient to reject non-flow correlations from dijets in $pp$ collisions, and methods with three or more subevents are required to suppress the non-flow contribution over the measured $\nch$ range. 

The middle row of Figure~\ref{fig:b2} shows $\scb$ from $\pPb$ collisions. Significant differences are observed between the standard method and the subevent methods over the full $\nch$ range. However, no differences are observed among the various subevent methods. These results suggest that the standard method is contaminated by large contributions from non-flow correlations at low $\nch$, and these contributions may not vanish even at large $\nch$ values. All subevent methods suggest an increase of $\scb$ toward lower $\nch$ for $\nch<40$, which may reflect some residual non-flow correlations in this region.

The bottom row of Figure~\ref{fig:b2} shows $\scb$ from Pb+Pb collisions. The $\scb$ values increase gradually with $\nch$ 
for all four methods. This increase reflects the known fact that the $v_2$ increases with $\nch$ in Pb+Pb collisions~\cite{Aad:2012bu}. The values from the standard method are systematically larger than those from the subevent methods, and this difference varies slowly with $\nch$, similar to the behaviour observed in $\pPb$ collisions in the high $\nch$ region.

\begin{figure}[h!]
\begin{center}
\includegraphics[width=0.8\linewidth]{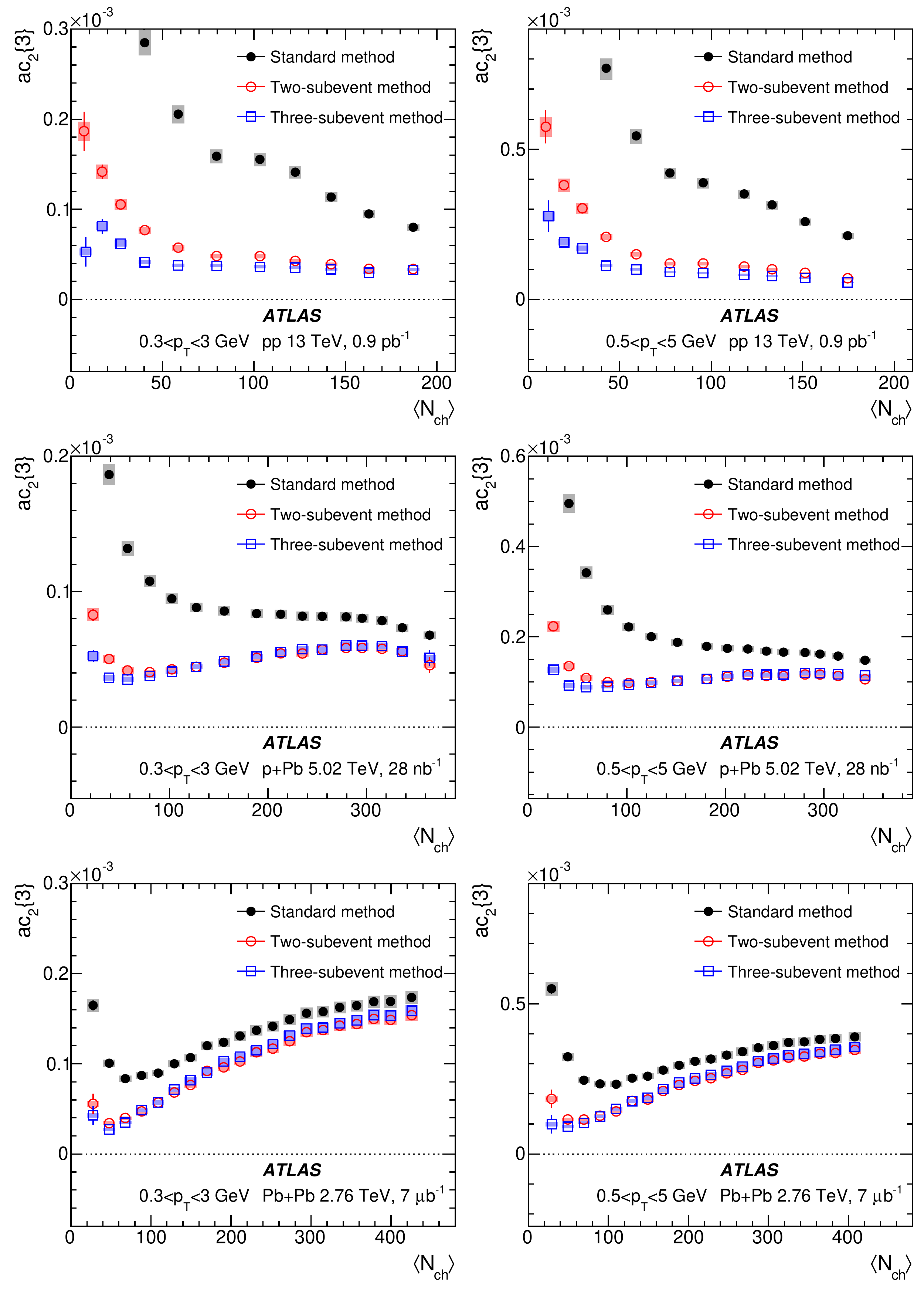}
\end{center}
\caption{\label{fig:b3} The asymmetric cumulant $\aca$ as a function of $\nch$ for $0.3<\pT<3$~GeV (left panels) and $0.5<\pT<5$~GeV (right panels) obtained for $pp$ collisions (top row), $\pPb$ collisions (middle row) and low-multiplicity Pb+Pb collisions (bottom row). In each panel, the $\aca$ is obtained from the standard method (filled symbol), two-subevent method (open circles), and three-subevent method (open squares). The error bars and shaded boxes represent the statistical and systematic uncertainties, respectively.}
\end{figure}

The results for the asymmetric cumulant $\aca$ are presented in Figure~\ref{fig:b3}. The top row shows the results obtained from the standard, two-subevent, and three-subevent methods from $pp$ collisions in $0.3<\pT<3$~GeV (left panel) and $0.5<\pT<5$~GeV (right panel). The results are positive for all methods. The results from the standard method are much larger than those from the subevent methods, consistent with the expectation that the standard method is more affected by non-flow correlations from dijets. Significant differences are also observed between the two-subevent and three-subevent methods at low $\nch$, but these differences decrease and disappear at $\nch>100$. The $\aca$ values from the three-subevent method show a slight increase for $\nch<40$ but are nearly constant for $\nch>40$. This behaviour suggests that in the three-subevent method, the non-flow contribution may play some role at $\nch<40$, but is negligible for $\nch>40$. Therefore, the $\aca$ from the three-subevent method supports the existence of a three-particle long-range collective flow that is nearly independent of $\nch$ in $pp$ collisions, consistent with the $\nch$-independent behaviour of $v_2$ and $v_4$ observed previously in the two-particle correlation analysis~\cite{Aaboud:2016yar}.

The middle and bottom rows of Figure~\ref{fig:b3} show $\aca$ from $\pPb$ and Pb+Pb collisions, respectively. The $\aca$ values from the standard method have a significant non-flow contribution up to $\nch \sim 200$ in $\pPb$ collisions and $\nch \sim 80$ in Pb+Pb collisions. In the subevent methods, the influence of non-flow contributions is very small for $\nch>60$ in both collision systems, and therefore the $\nch$ dependence of $\aca$ reflects the $\nch$ dependence of the $v_2$ and $v_4$. The $\aca$ values from the subevent methods increase with $\nch$, and the increase is stronger in Pb+Pb collisions. This is consistent with previous observations that $v_2$ and $v_4$ increase with $\nch$ more strongly in Pb+Pb than in $\pPb$ collisions~\cite{Aaboud:2017acw}.

The values of $\scb$ and $\aca$, which are both measures of correlations between $v_2$ and $v_4$, show significant differences between the standard method and the subevent methods, as shown in Figures~\ref{fig:b2} and \ref{fig:b3}. The $\nch$ dependence of these differences decreases gradually with $\nch$, and is consistent with an influence of non-flow that is expected to scale as $1/\nch$. However, these differences seem to persist for $\nch>200$ in $\pPb$ collisions and for $\nch>150$ in Pb+Pb collisions, which is not compatible with the predicted behaviour of non-flow correlations. The differences at large $\nch$ may arise from longitudinal flow decorrelations~\cite{Bozek:2010vz,Pang:2014pxa}, which have been measured by CMS~\cite{Khachatryan:2015oea} and ATLAS~\cite{Aaboud:2017tql}. Decorrelation effects are found to be large for $v_4$ and strongly correlated with $v_2$, and therefore they are expected to reduce the $\scb$ and $\aca$ in the subevent method. Therefore, the observed differences between the standard method and subevent method reflect the combined contribution from non-flow correlations, which dominates in the low $\nch$ region, and decorrelation, which is more important at large $\nch$ (see further discussion in the Appendix~\ref{sec:app2}). 

The results presented above suggest that the three-subevent method is sufficient to suppress most of the non-flow effects. It is therefore used as the default method for the discussion below.

\subsection{Comparison between collision systems}
\label{sec:3}

\begin{figure}[h!]
\begin{center}
\includegraphics[width=1\linewidth]{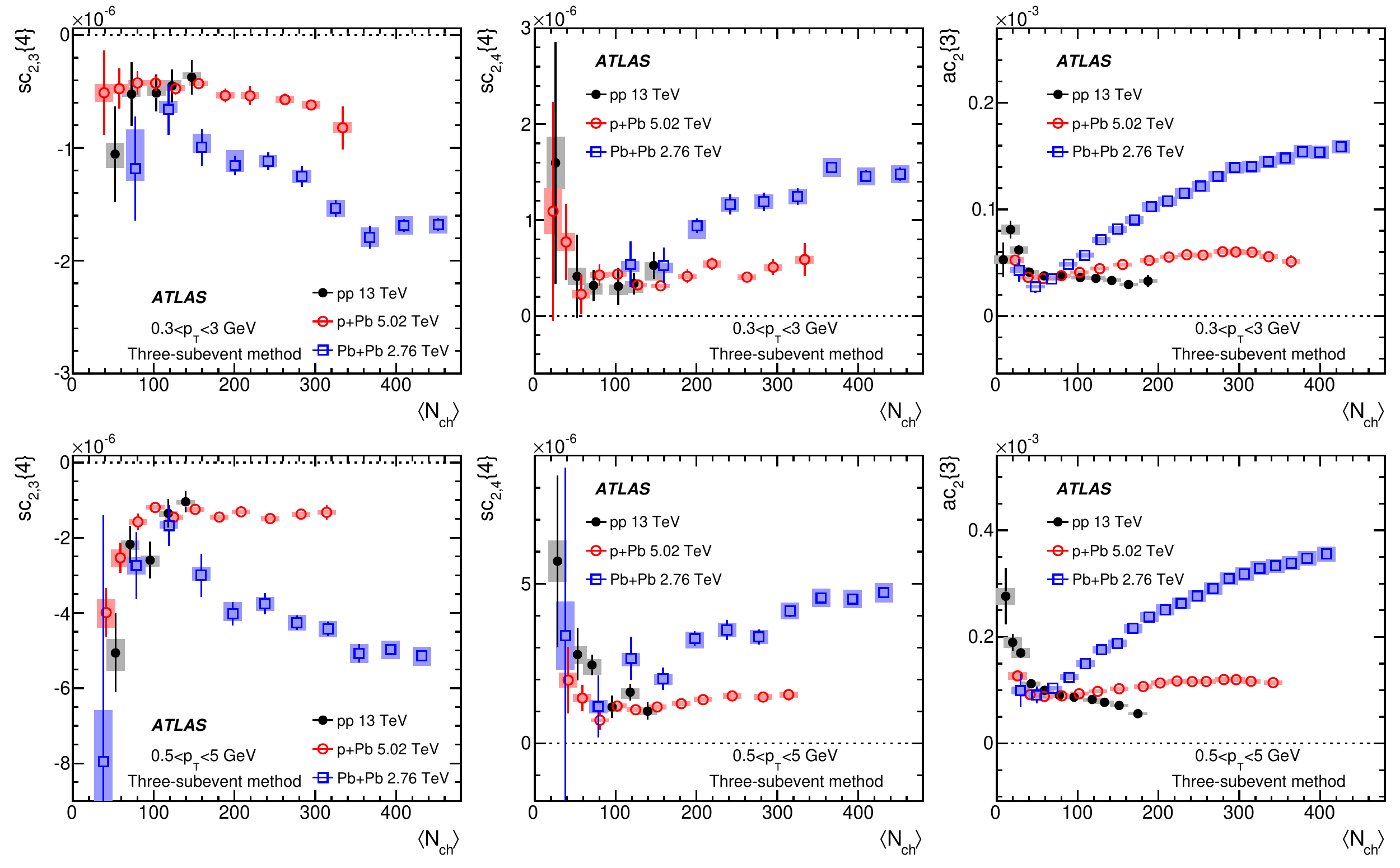}
\end{center}
\caption{\label{fig:c1} The $\nch$ dependence of $\sca$ (left panels), $\scb$ (middle panels) and $\aca$ (right panels) in $0.3<\pT<3$~GeV (top row) and $0.5<\pT<5$~GeV (bottom row) obtained for $pp$ collisions (solid circles), $\pPb$ collisions (open circles) and low-multiplicity Pb+Pb collisions (open squares). The error bars and shaded boxes represent the statistical and systematic uncertainties, respectively.}
\end{figure}

Figure~\ref{fig:c1} shows a direct comparison of cumulants for the three collision systems. The three panels in the top row show the results for $\sca$, $\scb$ and $\aca$, respectively, for $0.3<\pT<3$ GeV. These results support the existence of a negative correlation between $v_2$ and $v_3$ and a positive correlation between $v_2$ and $v_4$. Such correlation patterns have previously been observed in large collision systems~\cite{Aad:2014fla,Aad:2015lwa,ALICE:2016kpq}, but are now confirmed also in the small collision systems, once non-flow effects are adequately suppressed. In the multiplicity range covered by the $pp$ collisions, $\nch<150$, the results for symmetric cumulants $\sca$ and $\scb$ are similar among the three systems. In the range $\nch>150$, $|\sca|$ and $\scb$ are larger in Pb+Pb than in $\pPb$ collisions. The results for $\aca$ are similar among the three systems at $\nch<100$, but they deviate from each other at higher $\nch$. The $pp$ data are approximately constant or decrease slightly with $\nch$, while the $\pPb$ and Pb+Pb data show significant increases as a function of $\nch$. The bottom row shows the results for the higher $\pT$ range of $0.5<\pT<5$ GeV, where similar trends are observed. 

\begin{figure}[h!]
\begin{center}
\includegraphics[width=1\linewidth]{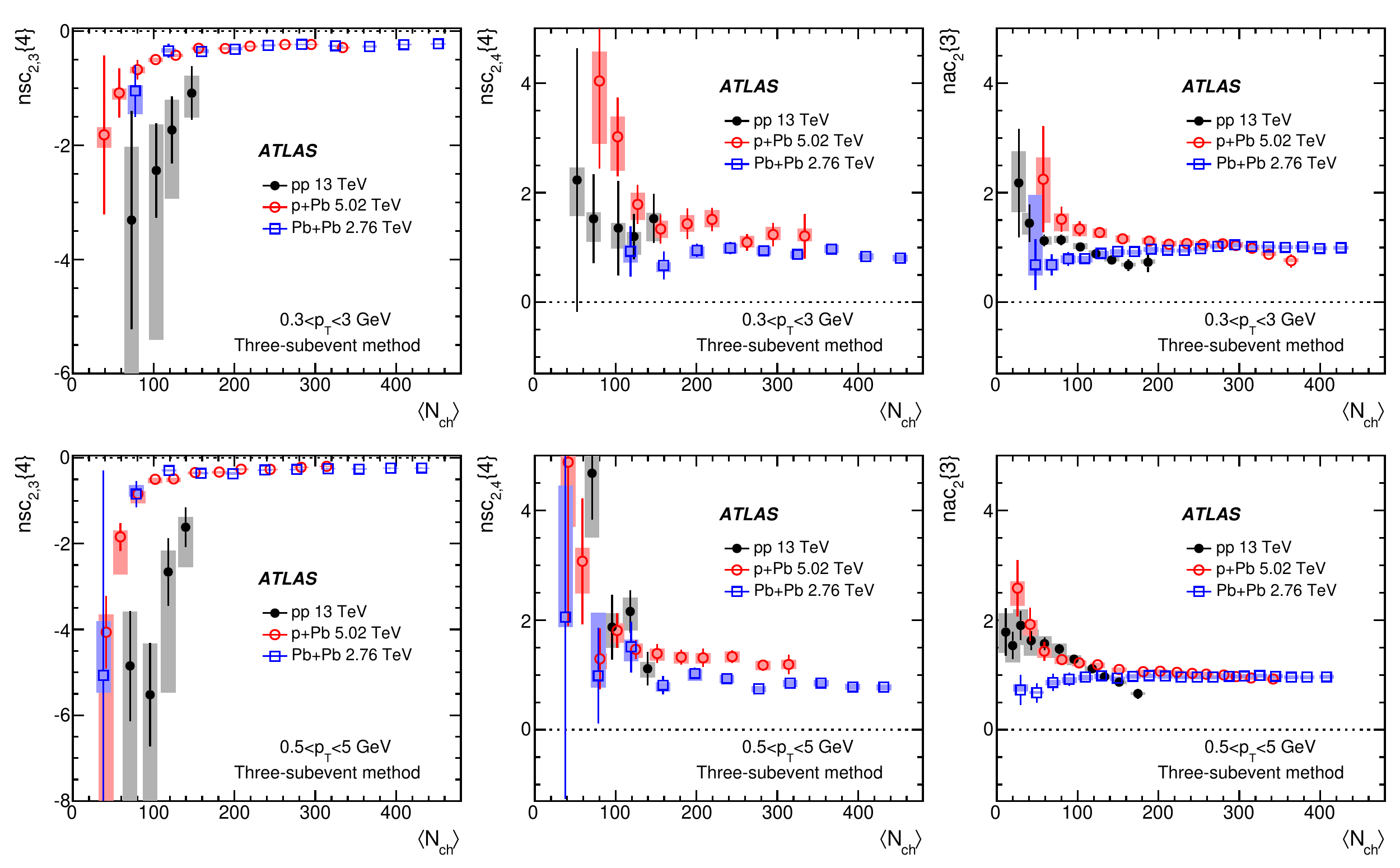}
\end{center}
\caption{\label{fig:c2} The $\nch$ dependence of $\nsca$ (left panels), $\nscb$ (middle panels) and $\naca$ (right panels) in $0.3<\pT<3$~GeV (top row) and $0.5<\pT<5$~GeV (bottom row) obtained for $pp$ collisions (solid circles), $\pPb$ collisions (open circles) and low-multiplicity Pb+Pb collisions (open squares). The error bars and shaded boxes represent the statistical and systematic uncertainties, respectively.}
\end{figure}

Figure~\ref{fig:c2} shows the results for normalised cumulants, $\nsca$, $\nscb$ and $\naca$, compared among the three systems. The normalised cumulants generally show a much weaker $\nch$ dependence at \mbox{$\nch>100$}, where the statistical uncertainties are small. This behaviour implies that the strong $\nch$ dependence of the $\scn$ and $\aca$ values reflects the $\nch$ dependence of the $v_n$ values, and these dependences are removed in the normalised cumulants. The normalised cumulants are also similar among different collision systems at large $\nch$, although some differences at the relative level of 20--30\% are observed for smaller $\nch$. The only exception is $\nsca$, whose values in the $pp$ collisions are very different from those in $\pPb$ and Pb+Pb collisions. In contrast, the $\sca$ values in Figure~\ref{fig:c1} are close among different systems. This suggests that the $\lr{v_3^2}$ values from the template fit method~\cite{Aaboud:2016yar} may be significantly underestimated. As pointed out in Ref.~\cite{Aaboud:2016yar} and emphasised in Appendix~\ref{sec:app1}, the template fit method, and other methods based on peripheral subtraction in general~\cite{Aad:2014lta,Khachatryan:2016txc}, tend to underestimate the odd flow harmonics, due to the presence of a large away-side peak at $\Dphi\sim\pi$ in the two-particle correlation function. The comparison of $\sca$ and $\nsca$ among different collision systems provides indirect evidence of this underestimation of $\lr{v_3^2}$.

Figure~\ref{fig:c2} shows that the normalised cumulants are consistent between $0.3<\pT<3$~GeV and $0.5<\pT<5$~GeV. On the other hand, the magnitudes of the cumulants in Figure~\ref{fig:c1} differ by a large factor between the two $\pT$ ranges: about a factor of three for $\sca$ and $\scb$, and a factor of two for $\aca$. These results suggest that the $\pT$ dependence of $\sca$, $\scb$ and $\aca$ largely reflects the $\pT$ dependence of the $v_n$ at the single-particle level.

\section{Discussion}
\label{sec:sum}
Three- and four-particle cumulants involving correlations between two harmonics of different order $v_n$ and $v_m$ are measured in $\sqrt{s}$ = 13~TeV $\pp$, $\sqrtsnn$ = 5.02~TeV $\pPb$, and low-multiplicity $\sqrtsnn$ = 2.76~TeV Pb+Pb collisions with the ATLAS detector at the LHC, with total integrated luminosities of 0.9~$\mathrm{pb}^{-1}$, 28~$\mathrm{nb}^{-1}$, and 7~$\mu\mathrm{b}^{-1}$, respectively. The correlation between $v_n$ and $v_m$ is studied using four-particle symmetric cumulants, $\sca$ and $\scb$, and the three-particle asymmetric cumulant $\aca$. The symmetric cumulants $\scn= \lr{v_n^2v_m^2}-\lr{v_n^2}\lr{v_m^2}$ probe the correlation of the flow magnitudes, while the asymmetric cumulant \mbox{$\mathrm{ac}_2\{3\}= \lr{v_2^2v_{4}\cos4(\Phi_2-\Phi_{4})}$} is sensitive to correlations involving both the flow magnitude $v_n$ and flow phase $\Phi_n$. They are calculated using the standard cumulant method, as well as the two-, three- and four-subevent methods to suppress non-flow effects. The final results are presented as a function of the average number of charged particles with $\pT > 0.4$~GeV, $\nch$.

Significant differences are observed between the standard method and the subevent methods over the full $\nch$ range in $pp$ collisions, as well as over the low $\nch$ range in $\pPb$ and Pb+Pb collisions. The differences are larger for particles at higher $\pT$ or at smaller $\nch$. When analysed with the standard method in $pp$ collisions, this behaviour is compatible with the dominance of the non-flow correlations rather than the long-range collective flow correlations. Systematic, but much smaller, differences are also observed in the low $\nch$ region between the two-subevent method and three- or four-subevent methods, which indicate that the two-subevent method may still be affected by correlations arising from jets. On the other hand no differences are observed between the three-subevent and four-subevent methods, within experimental uncertainties, suggesting that methods with three or more subevents are sufficient to reject non-flow correlations from jets. Therefore, the three-subevent method is used to present the main results in this analysis.

The three-subevent method provides a measurement of negative $\sca$ and positive $\scb$ and $\aca$ over nearly the full $\nch$ range and in all three collision systems. These results indicate a negative correlation between $v_2$ and $v_3$ and a positive correlation between $v_2$ and $v_4$. Such correlation patterns have previously been observed in large collision systems~\cite{Aad:2014fla,Aad:2015lwa,ALICE:2016kpq}, but are now confirmed in small collision systems, once non-flow effects are adequately suppressed. The values of $\sca$ and $\scb$ are consistent in $pp$ and $\pPb$ collisions over the same $\nch$ range, but their magnitudes at large $\nch$ are much smaller than those for Pb+Pb collisions. The values of $\aca$ are similar at very low $\nch$ among the three systems, but are very different at large $\nch$. On the other hand, after scaling by the $\lr{v_n^2}$ estimated from a two-particle analysis~\cite{Aad:2015gqa,Aaboud:2016yar}, the resulting normalised cumulants $\nsca$, $\nscb$ and $\naca$ show a much weaker dependence on $\nch$, and their values are much closer to each other among the three systems. The magnitudes of the normalised cumulants are also similar to each other for $0.5<\pT<5$~GeV as well as $0.3<\pT<3$~GeV. This suggests that the $\nch$, $\pT$ and system dependence of the $\sca$, $\scb$ and $\aca$ reflect mostly the $\nch$, $\pT$ and system dependence of $\lr{v_n^2}$, but the relative strengths of the correlations are similar for the three collision systems. 

The new results obtained with the subevent cumulant technique provide further evidence that the ridge is indeed a long-range collective phenomenon involving many particles distributed across a broad rapidity interval. The similarity between different collision systems for $\nsca$, $\nscb$ and $\naca$, and the weak dependence of these observables on the $\pT$ range and $\nch$, largely free from non-flow effects, provide an important input towards understanding the space-time dynamics and the properties of the medium created in small collision systems. These results provide inputs to distinguish between models based on initial-state momentum correlations and models based on final-state hydrodynamics.

\clearpage
\appendix
\part*{Appendix}
\addcontentsline{toc}{part}{Appendix}
\section{Improvement to the template fit procedure}
\label{sec:app1}
In order to separate the long-range ridge from other non-flow sources, especially dijets, the ATLAS Collaboration developed a template fitting procedure described in Refs.~\cite{Aad:2015gqa,Aaboud:2016yar}.  The first step is to construct a $\Dphi$ distribution of particle pairs with large pseudorapidity separation $|\Deta|>2$, the so-called ``per-trigger'' particle yield, $Y(\Dphi)$, for a given $\nchrec$ range. The $|\Deta|>2$ requirement suppresses the intra-jet and other short-range correlations, and in small collision systems the resulting $Y(\Dphi)$ distributions are known to be dominated by away-side jet correlations~\cite{Aad:2012gla,Aad:2014lta,Aad:2015gqa}. This away-side non-flow component is peaked at $\Dphi\sim\pi$, and leads to a significant bias in the flow coefficients $v_n$, especially for the odd harmonics. 

To subtract the away-side jet correlations, the measured $Y(\Dphi)$ distribution in a given $\nchrec$ interval is assumed to be a sum of a scaled ``peripheral'' distribution $Y(\Dphi)^{\mathrm{peri}}$, obtained for low-multiplicity events $\nchrec<20$, and a constant pedestal modulated by $\cos(n\Dphi)$ for $n\geq2$~\cite{Aad:2015gqa,Aaboud:2016yar}:
\begin{eqnarray}\label{eq:a1}
Y(\Dphi) = F\;Y(\Dphi)^{\mathrm{peri}} + G^{\mathrm{tmp}} \left(1+2\sum_{n=2}^{\infty} v_n\{2,\mathrm{tmp}\}^2 \cos n\Dphi\right).
\end{eqnarray}
The scale factor $F$ and pedestal $G^{\mathrm{tmp}}$ are fixed by the fit, and $v_n\{2,\mathrm{tmp}\}$ are calculated from a Fourier transform. On the other hand, both $Y(\Dphi)$ and $Y(\Dphi)^{\mathrm{peri}}$ contain a dijet component and flow component:
\begin{eqnarray}\label{eq:a2a}
Y(\Dphi) &=&Y(\Dphi)_{\mathrm{jet}}^{\mathrm{cent}} + G^{\mathrm{cent}} \left(1+2\sum_{n=2}^{\infty} v_n\{2\}^2 \cos n\Dphi\right)\;,\\\label{eq:a2b}
Y(\Dphi)^{\mathrm{peri}} &=&Y(\Dphi)_{\mathrm{jet}}^{\mathrm{peri}} + G^{\mathrm{peri}} \left(1+2\sum_{n=2}^{\infty} v_n\{2,\mathrm{peri}\}^2 \cos n\Dphi\right).
\end{eqnarray}
With the assumption that the shape of the dijet component is independent of $\nchrec$, and the magnitudes of the dijet components  are related by the scale factor $F$: $Y(\Dphi)_{\mathrm{jet}}^{\mathrm{cent}}=FY(\Dphi)_{\mathrm{jet}}^{\mathrm{peri}}$, Eq.~\eqref{eq:a1} can be written as:
\begin{eqnarray}\nonumber
Y(\Dphi) =Y(\Dphi)_{\mathrm{jet}}^{\mathrm{cent}} + (G^{\mathrm{tmp}}+FG^{\mathrm{peri}})+ 2\sum_{n=2}^{\infty}\left(G^{\mathrm{tmp}}v_n\{2,\mathrm{tmp}\}^2+FG^{\mathrm{peri}}v_n\{2,\mathrm{peri}\}^2\right) \cos n\Dphi.
\end{eqnarray}
Comparing with Eqs.~\eqref{eq:a2a} and \eqref{eq:a2b}, one obtains $G^{\mathrm{cent}} = G^{\mathrm{tmp}}+FG^{\mathrm{peri}}$ and the following relation:
\begin{eqnarray}\nonumber
v_n\{2\}^2 = v_n\{2,\mathrm{tmp}\}^2 -\frac{F\;G^{\mathrm{peri}}}{G^{\mathrm{cent}}}\left(v_n\{2,\mathrm{tmp}\}^2-v_n\{2,\mathrm{peri}\}^2\right)\;,
\end{eqnarray}
which shows that $v_n\{2,\mathrm{tmp}\}$ from the template fit differs from the true $v_n\{2\}$ by a correction term that vanishes if and only if $v_n\{2\}$ is independent of $\nchrec$. Since the true flow harmonics in the peripheral interval $v_n\{2,\mathrm{peri}\}$ are unknown in principle, the correction is applied starting from the third-lowest $\nchrec$ interval ($40\leq\nchrec<60$) in this analysis, by using $v_n\{2,\mathrm{tmp}\}$ of the second $\nchrec$ interval ($20\leq\nchrec<40$) as an estimate of the true flow harmonics. Since the non-flow contribution primarily affects the odd harmonics, the $v_3\{2,\mathrm{tmp}\}^2$ may become negative in the first few $\nchrec$ intervals in $pp$ collisions. In such cases, the correction starts from the second $\nchrec$ interval with positive $v_3\{2,\mathrm{tmp}\}^2$ ($60\leq\nchrec<80$) by using $v_3\{2,\mathrm{tmp}\}$ from the previous $\nchrec$ interval ($40\leq\nchrec<60$).

\begin{figure}[h!]
\begin{center}
\includegraphics[width=1\linewidth]{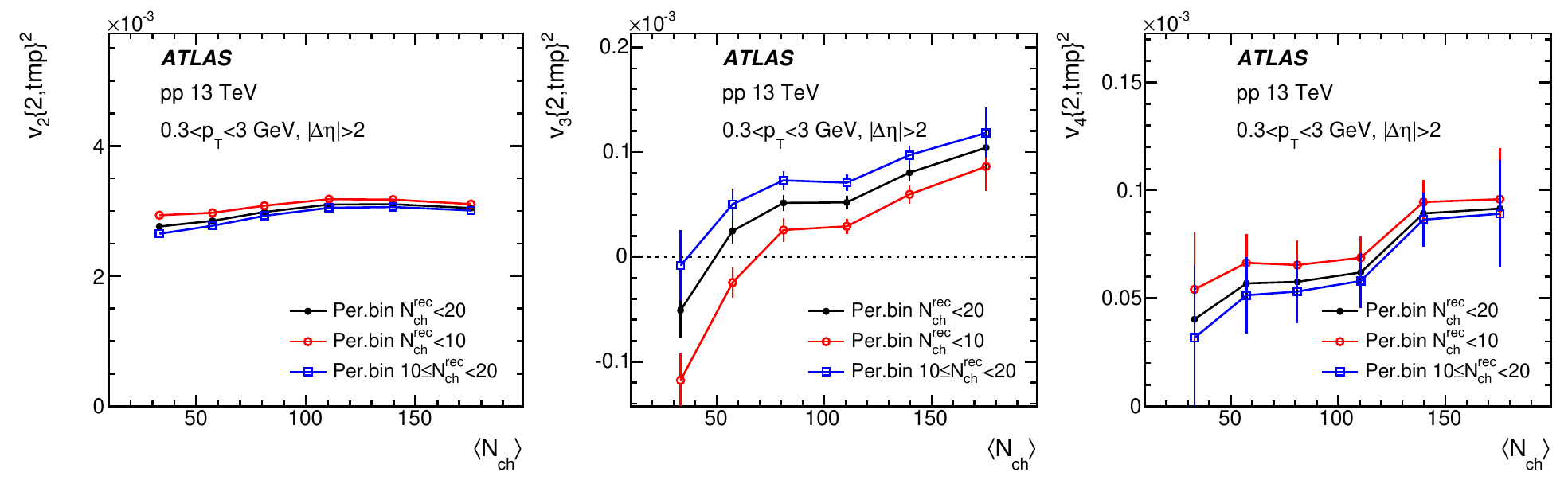}
\end{center}
\caption{\label{fig:app0} The values of $v_n\{2,\mathrm{tmp}\}^2$ obtained following the template fit procedure given in Eq.~\eqref{eq:a1}~\cite{Aaboud:2016yar} in $pp$ collisions for $n=2$ (left panel), $n=3$ (middle panel) and $n=4$ (right panel). In each panel, the values are calculated for three peripheral $\nchrec$ intervals: $\nchrec<20$, $\nchrec<10$ and $10\leq\nchrec<20$. Only statistical uncertainties are shown.}
\end{figure}

One important feature of the template fit analysis is the assumption that the dijet component $Y(\Dphi)_{\mathrm{jet}}$ is independent of $\nch$. In Ref.~\cite{Aaboud:2016yar}, the uncertainty associated with this assumption is studied by changing the default peripheral interval from $\nchrec<20$ to $\nchrec<10$ and $10\leq\nchrec<20$. It was found that the $v_n\{2,\mathrm{tmp}\}$ values are relatively insensitive to the choice of peripheral interval for $n=2$ and $n=4$, but the sensitivity is much larger for $n=3$. This finding is reproduced in Figure~\ref{fig:app0} for $\pp$ collisions, which shows that the $v_3\{2,\mathrm{tmp}\}^2$ values obtained via Eq.~\eqref{eq:a1} differ substantially for the different $\nchrec$ ranges.

In addition to the template fit with and without the above mentioned correction procedure, the ATLAS and CMS collaborations also calculated directly the $v_n\{2\}$ values via a Fourier transform of the $Y(\Dphi)$ distribution without dijet subtraction~\cite{Aaboud:2016yar,Sirunyan:2017uyl}.  The differences between the direct Fourier transform and template fit reflect mainly the away-side jet contribution subtracted by the template fit procedure, and therefore give a sense of the magnitude of unknown systematic uncertainties associated with the template fit procedure. If these differences are too large, the $v_n\{2,\mathrm{tmp}\}$ values may be sensitive to the systematic effects associated with the assumption that the shape of $Y(\Dphi)_{\mathrm{jet}}$ is independent of $\nchrec$. 

\begin{figure}[h!]
\begin{center}
\includegraphics[width=1\linewidth]{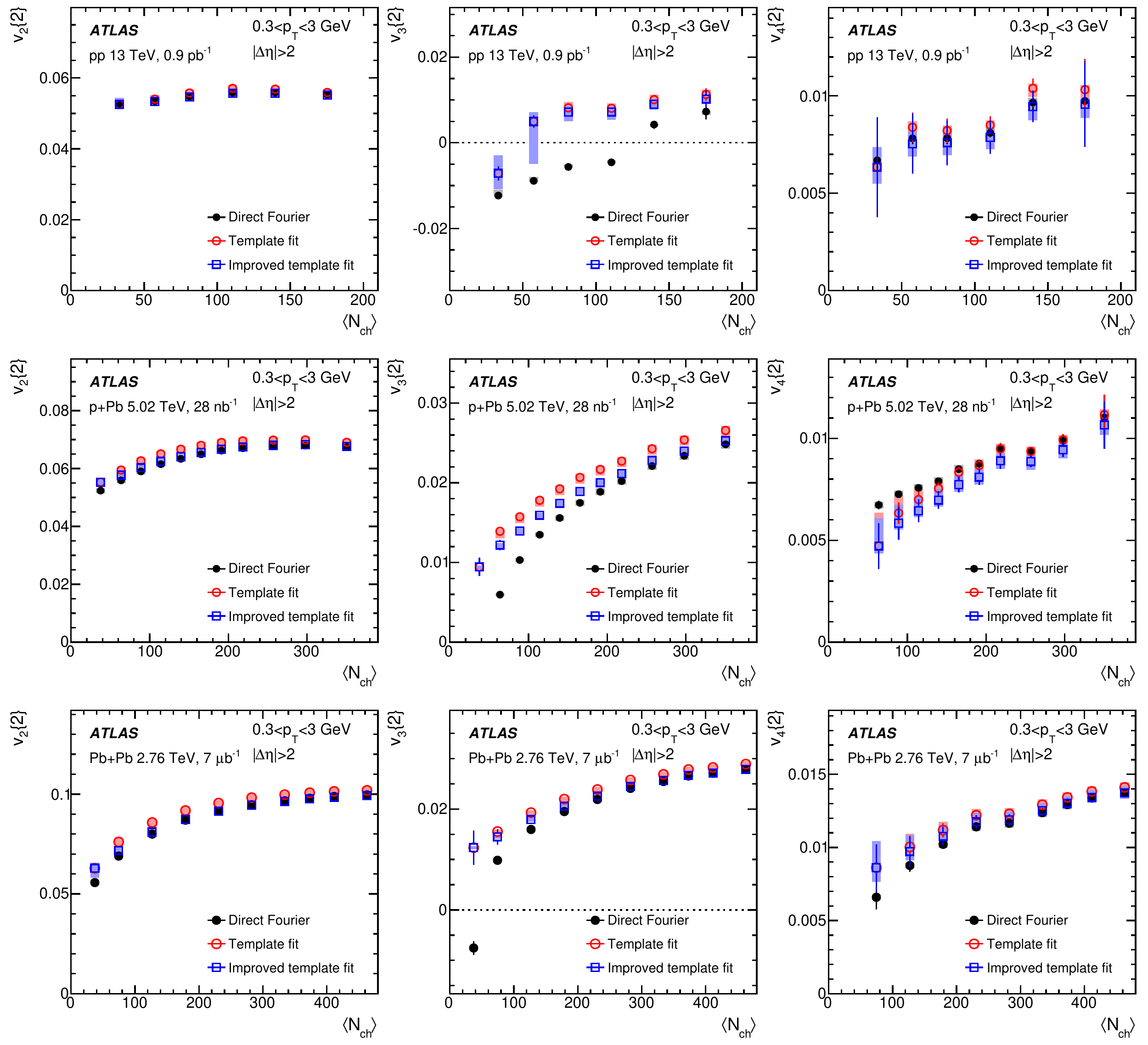}
\end{center}
\caption{\label{fig:app1} The $v_2$ (left column), $v_3$ (middle column) and $v_4$ (right column) obtained from two-particle correlations in $0.3<\pT<3$ GeV in $pp$ (top row), $\pPb$ (middle row) and Pb+Pb (bottom row) collisions. In each panel, they are compared between three methods: direct Fourier transformation (solid circles), template fit (open circles) and the improved template fit (open squares). The error bars and shaded boxes represent the statistical and systematic uncertainties, respectively.}
\end{figure}

Figure~\ref{fig:app1} compares the $v_n\{2\}$ in $0.3<\pT<3$ GeV obtained from $Y(\Dphi)$ using three methods: a direct Fourier transform (solid circles), a template fit (open circles) and a template fit corrected for the bias (open squares), as described above. The systematic uncertainties for the template fit results are nearly the same as those from Ref.~\cite{Aaboud:2016yar}. Figure~\ref{fig:app1} shows that the changes introduced by the correction procedure described above are small in all cases and for all harmonics. The values of the even-order harmonics, $v_2$ and $v_4$, are also quite similar to those obtained from the direct Fourier transformation, reflecting the fact that the dijet correlations have very little influence on the even-order harmonics.  On the other hand, significant differences are observed between the direct Fourier transform and template fit for $v_3$, especially in the $pp$ collisions, due to the influence of $Y(\Dphi)_{\mathrm{jet}}$, a trend observed and discussed previously in Refs.~\cite{Aaboud:2016yar,Khachatryan:2016txc}. The template fit procedure is able to subtract the dijet correlations and change the sign of $v_3$, but also introduces a large uncertainty associated with the procedure. As discussed in Section~\ref{sec:3}, the behaviour of the symmetric cumulants $\sca$ in Figure~\ref{fig:c1} and normalised cumulants $\nsca$ in Figure~\ref{fig:c2} in $pp$ collisions, suggest that the $v_3$ values from the template fit procedure are significantly underestimated due to the presence of a large residual non-flow bias. In contrast, the differences of $v_3$ between the direct Fourier transform and the template fit are much smaller in the $\pPb$ and the Pb+Pb collisions, except in the very low $\nch$ region. Therefore, the $v_3$ values in $\pPb$ and Pb+Pb systems extracted from the template fit procedure are expected to be less affected by the dijets.

\section{Effects of flow decorrelations in the subevent cumulant methods}
\label{sec:app2}
As discussed in Section~\ref{sec:result}, the differences between the standard method and subevent methods for $\aca$ can be partially attributed to longitudinal flow decorrelations~\cite{Bozek:2010vz,Aaboud:2017tql}. Since subevent methods correlate flow vectors obtained from different $\eta$ regions, the influence of decorrelations can be studied by comparing different variants of the three-subevent method. This comparison is carried out using asymmetric cumulant $\mathrm{ac}_2=\lr{{\bm V}_2^2{\bm V}_{4}^*}$, however, the statistical precision of $\scn$ is not sufficient for such comparison.

\begin{figure}[h!]
\begin{center}
\includegraphics[width=1\linewidth]{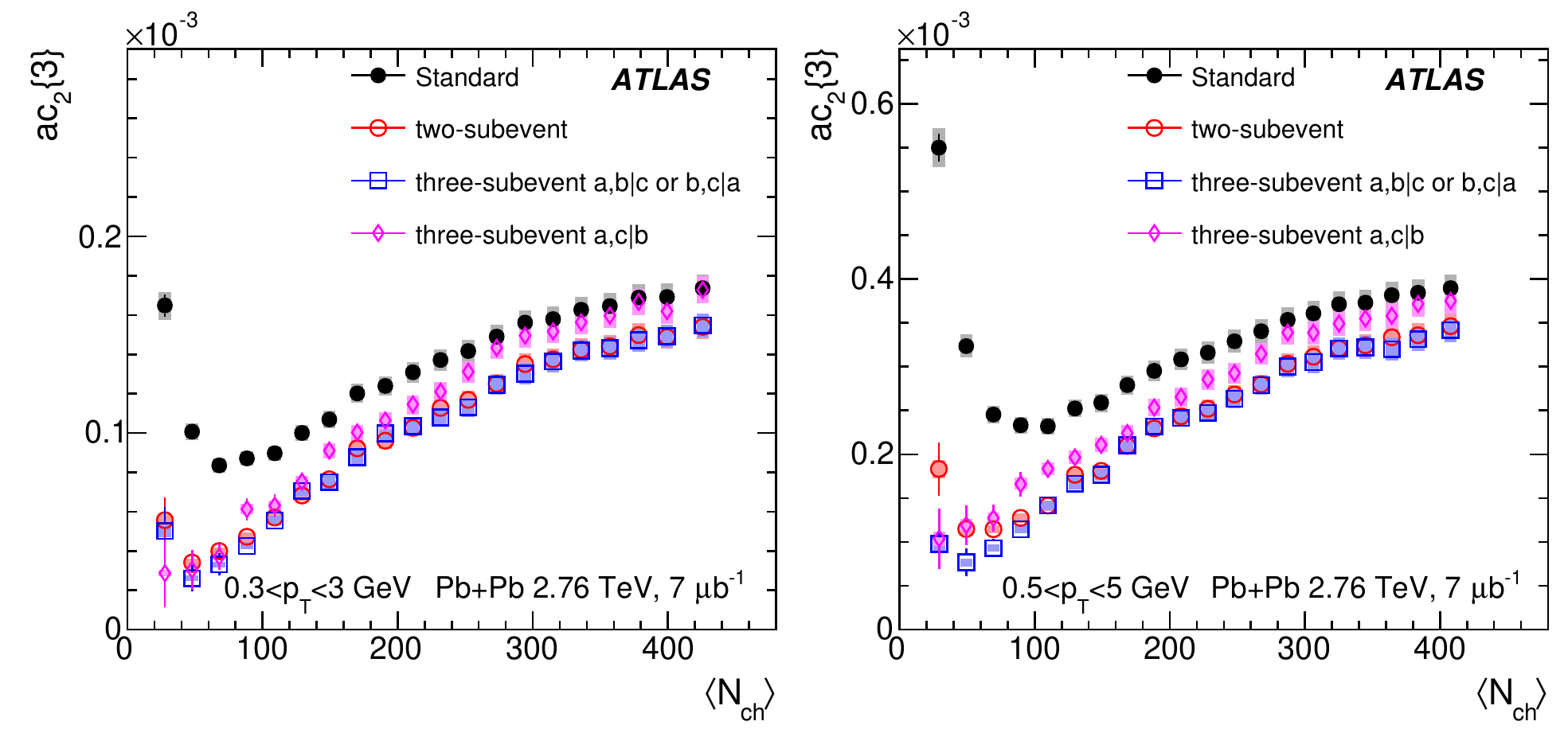}
\end{center}
\caption{\label{fig:app2} The $\aca$ in $0.3<\pT<3$ GeV (left panel) and $0.5<\pT<5$ GeV (right panel) in Pb+Pb collisions. In each panel, they are compared between the standard method (solid circles), two-subevent method (open circles), three-subevent where ${\bm V}_{4}$ is determined in subevent $a$ or $c$ (open boxes), and three-subevent where ${\bm V}_{4}$ is determined in subevent $b$ (diamonds) according to Eq.~\eqref{eq:6c}. The error bars and shaded boxes represent the statistical and systematic uncertainties, respectively.}
\end{figure}

The three-subevent method uses flow vectors from three subevents $a$, $b$ and $c$, covering $-2.5<\eta_a<-2.5/3$, $|\eta_b|\leq2.5/3$ and $2.5/3<\eta_c<2.5$, and have three independent definitions for asymmetric cumulant: $\mathrm{ac}_2^{a,b|c}\{3\}$, $\mathrm{ac}_n^{b,c|a}\{3\}$ and $\mathrm{ac}_2^{a,c|b}\{3\}$. Because of symmetry between subevents $a$ and $c$, $\mathrm{ac}_2^{a,b|c}\{3\}$ and $\mathrm{ac}_n^{b,c|a}\{3\}$ measure the same physics, and are therefore averaged into a single result, denoted as $\mathrm{ac}_2^{a,b|c \mathrm {\;or\;} b,c|a}\{3\}$. Figure~\ref{fig:app2} compares the two three-subevent results with results for the standard and two-subevent methods. The results based on various subevent methods show a small decrease with $\nch$ in the $\nch<50$ region, reflecting a modest contribution from non-flow. On the other hand, all the subevent-based results increase gradually with $\nch$ for $\nch>50$, reflecting a dominant contribution from flow. 

\begin{figure}[h!]
\begin{center}
\includegraphics[width=1\linewidth]{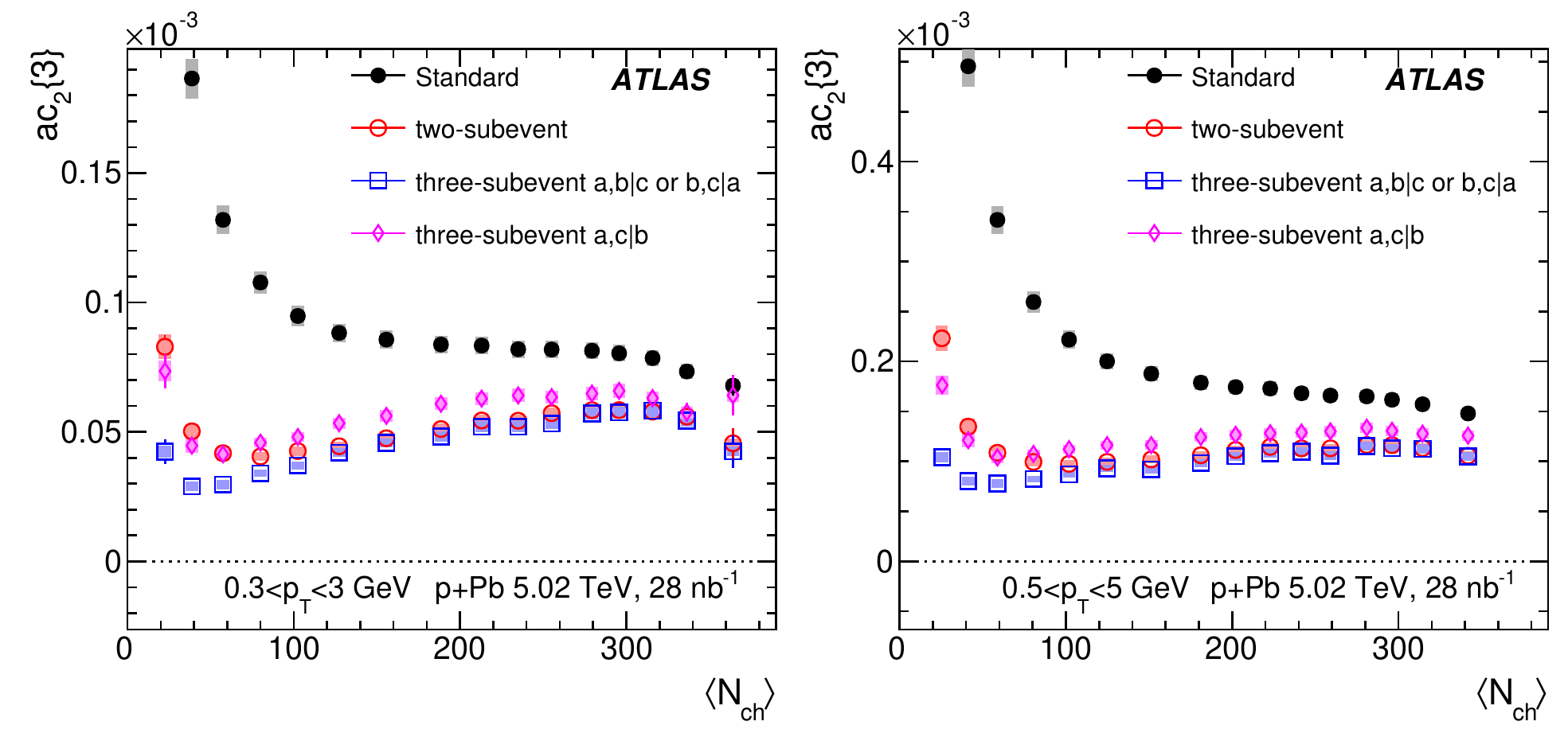}
\end{center}
\caption{\label{fig:app3} The $\aca$ in $0.3<\pT<3$ GeV (left panel) and $0.5<\pT<5$ GeV (right panel) in $\pPb$ collisions. In each panel, they are compared between standard method (solid circles), two-subevent method (open circles), three-subevent where ${\bm V}_{4}$ is determined in subevent $a$ or $c$ (open boxes), and three-subevent where ${\bm V}_{4}$ is determined in subevent $b$ (diamonds) according to Eq.~\eqref{eq:6c}. The error bars and shaded boxes represent the statistical and systematic uncertainties, respectively.}
\end{figure}

\begin{figure}[h!]
\begin{center}
\includegraphics[width=1\linewidth]{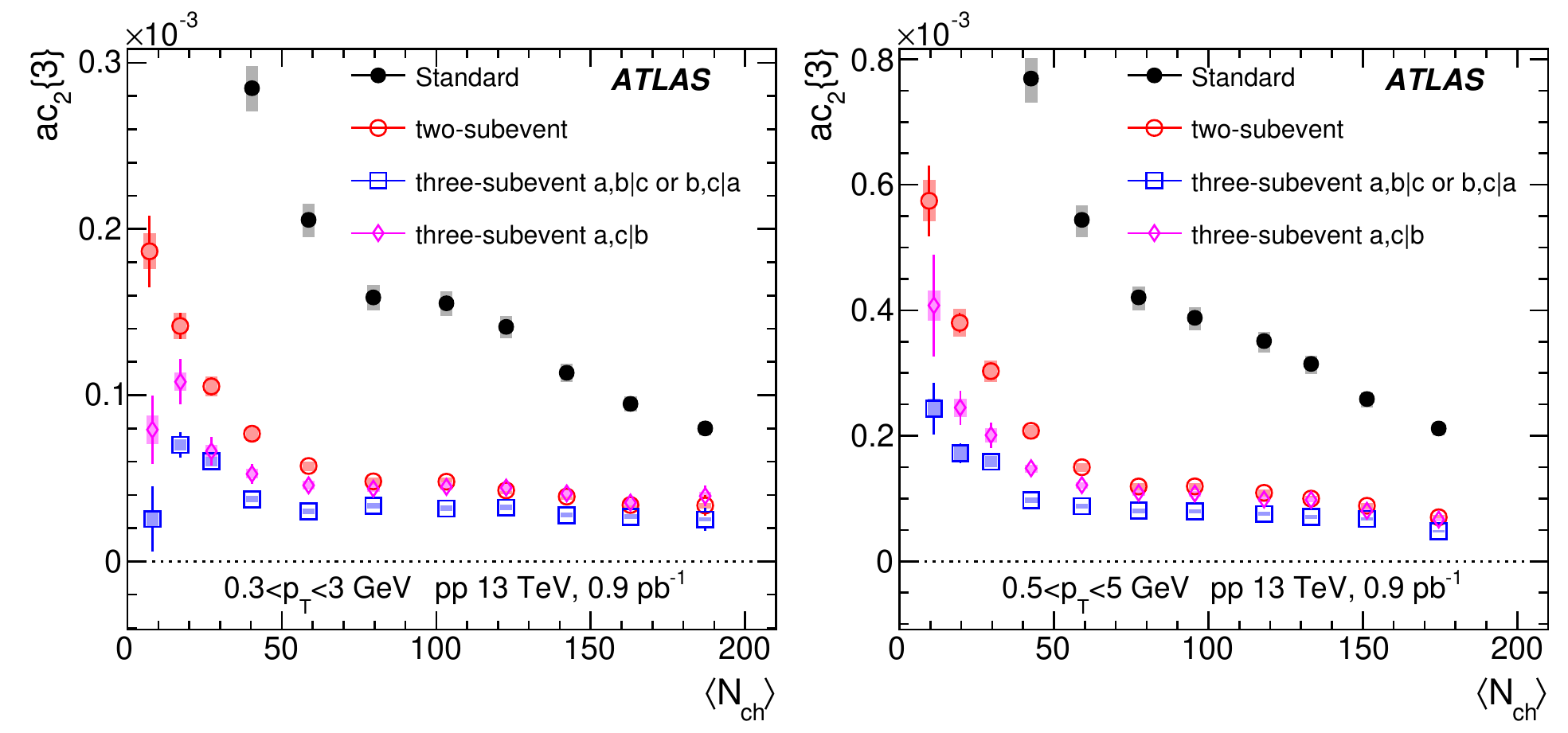}
\end{center}
\caption{\label{fig:app4} The $\aca$ in $0.3<\pT<3$ GeV (left panel) and $0.5<\pT<5$ GeV (right panel) in $pp$ collisions. In each panel, they are compared between standard method (solid circles), two-subevent method (open circles), three-subevent where ${\bm V}_{4}$ is determined in subevent $a$ or $c$ (open boxes), and three-subevent where ${\bm V}_{4}$ is determined in subevent $b$ (diamonds) according to Eq.~\eqref{eq:6c}. The error bars and shaded boxes represent the statistical and systematic uncertainties, respectively.}
\end{figure}

Figure~\ref{fig:app2} shows that the values of $\mathrm{ac}_2^{a,c|b}\{3\}=\lr{{\bm V}_{2,a}{\bm V}_{4,b}^*{\bm V}_{2,c}}$ are larger than $\mathrm{ac}_2^{a,b|c \mathrm{\;or\;} b,c|a}\{3\}$ at larger $\nch$ region. This is because the subevent for ${\bm V}_{4}$ is in between the two subevents used to calculate ${\bm V}_{2}$. This configuration has much smaller decorrelation effects~\cite{Aaboud:2017tql}. For $\mathrm{ac}_2^{a,b|c \mathrm {\;or\;} b,c|a}\{3\}$, the two subevents for ${\bm V}_{2}$ are on the same side of the subevent for ${\bm V}_{4}$, leading to larger decorrelation effects. Interestingly, such configuration gives results that are very similar to those from the two-subevent method. Figures~\ref{fig:app3} and \ref{fig:app4} show the results for the $\pPb$ and $pp$ collisions, respectively. Similar observations as in Pb+Pb collisions can be made, although in $pp$ collisions the results from the two-subevent method are larger than those obtained with the three-subevent method, due to significant non-flow contribution even in the large $\nch$ region.

\section*{Acknowledgements}

We thank CERN for the very successful operation of the LHC, as well as the
support staff from our institutions without whom ATLAS could not be
operated efficiently.

We acknowledge the support of ANPCyT, Argentina; YerPhI, Armenia; ARC, Australia; BMWFW and FWF, Austria; ANAS, Azerbaijan; SSTC, Belarus; CNPq and FAPESP, Brazil; NSERC, NRC and CFI, Canada; CERN; CONICYT, Chile; CAS, MOST and NSFC, China; COLCIENCIAS, Colombia; MSMT CR, MPO CR and VSC CR, Czech Republic; DNRF and DNSRC, Denmark; IN2P3-CNRS, CEA-DRF/IRFU, France; SRNSFG, Georgia; BMBF, HGF, and MPG, Germany; GSRT, Greece; RGC, Hong Kong SAR, China; ISF and Benoziyo Center, Israel; INFN, Italy; MEXT and JSPS, Japan; CNRST, Morocco; NWO, Netherlands; RCN, Norway; MNiSW and NCN, Poland; FCT, Portugal; MNE/IFA, Romania; MES of Russia and NRC KI, Russian Federation; JINR; MESTD, Serbia; MSSR, Slovakia; ARRS and MIZ\v{S}, Slovenia; DST/NRF, South Africa; MINECO, Spain; SRC and Wallenberg Foundation, Sweden; SERI, SNSF and Cantons of Bern and Geneva, Switzerland; MOST, Taiwan; TAEK, Turkey; STFC, United Kingdom; DOE and NSF, United States of America. In addition, individual groups and members have received support from BCKDF, CANARIE, CRC and Compute Canada, Canada; COST, ERC, ERDF, Horizon 2020, and Marie Sk{\l}odowska-Curie Actions, European Union; Investissements d' Avenir Labex and Idex, ANR, France; DFG and AvH Foundation, Germany; Herakleitos, Thales and Aristeia programmes co-financed by EU-ESF and the Greek NSRF, Greece; BSF-NSF and GIF, Israel; CERCA Programme Generalitat de Catalunya, Spain; The Royal Society and Leverhulme Trust, United Kingdom. 

The crucial computing support from all WLCG partners is acknowledged gratefully, in particular from CERN, the ATLAS Tier-1 facilities at TRIUMF (Canada), NDGF (Denmark, Norway, Sweden), CC-IN2P3 (France), KIT/GridKA (Germany), INFN-CNAF (Italy), NL-T1 (Netherlands), PIC (Spain), ASGC (Taiwan), RAL (UK) and BNL (USA), the Tier-2 facilities worldwide and large non-WLCG resource providers. Major contributors of computing resources are listed in Ref.~\cite{ATL-GEN-PUB-2016-002}.

\clearpage
\bibliographystyle{atlasBibStyleWithTitle}
\bibliography{hion-2017-07}

\providecommand{\href}[2]{#2}\begingroup\raggedright\begin{thebibliography}{10}

\bibitem{Shuryak:2014zxa}
E.~Shuryak, {\em {Strongly coupled quark-gluon plasma in heavy ion
  collisions}}, \href{http://dx.doi.org/10.1103/RevModPhys.89.035001}{Rev. Mod.
  Phys. {\bfseries 89} (2017) 035001},
\href{http://arxiv.org/abs/1412.8393}{{\ttfamily arXiv:1412.8393 [hep-ph]}}.

\bibitem{CMS:2012qk}
{CMS Collaboration}, {\em {Observation of long-range near-side angular
  correlations in proton-lead collisions at the LHC}},
  \href{http://dx.doi.org/10.1016/j.physletb.2012.11.025}{Phys.~Lett.~B
  {\bfseries 718} (2013) 795}, \href{http://arxiv.org/abs/1210.5482}{{\ttfamily
  arXiv:1210.5482 [nucl-ex]}}.

\bibitem{Abelev:2012ola}
{ALICE Collaboration}, {\em {Long-range angular correlations on the near and
  away side in $p$-Pb collisions at $\sqrt{s_{NN}}=5.02$ TeV}},
  \href{http://dx.doi.org/10.1016/j.physletb.2013.01.012}{Phys.~Lett.~B
  {\bfseries 719} (2013) 29}, \href{http://arxiv.org/abs/1212.2001}{{\ttfamily
  arXiv:1212.2001 [nucl-ex]}}.

\bibitem{Aad:2012gla}
{ATLAS Collaboration}, {\em {Observation of Associated Near-side and Away-side
  Long-range Correlations in $\sqrt{s_{NN}}$=5.02 TeV Proton-lead Collisions
  with the ATLAS Detector}},
  \href{http://dx.doi.org/10.1103/PhysRevLett.110.182302}{Phys.~Rev.~Lett.
  {\bfseries 110} (2013) 182302},
  \href{http://arxiv.org/abs/1212.5198}{{\ttfamily arXiv:1212.5198 [hep-ex]}}.

\bibitem{Aad:2014lta}
{ATLAS Collaboration}, {\em {Measurement of long-range pseudorapidity
  correlations and azimuthal harmonics in $\sqrt{s_{\mathrm{NN}}}$ = 5.02 TeV
  proton-lead collisions with the ATLAS detector}},
  \href{http://dx.doi.org/10.1103/PhysRevC.90.044906}{Phys.~Rev.~C {\bfseries
  90} (2014) 044906},
\href{http://arxiv.org/abs/1409.1792}{{\ttfamily arXiv:1409.1792 [hep-ex]}}.

\bibitem{Khachatryan:2015waa}
{CMS Collaboration}, {\em {Evidence for Collective Multiparticle Correlations
  in p-Pb Collisions}},
  \href{http://dx.doi.org/10.1103/PhysRevLett.115.012301}{Phys. Rev. Lett.
  {\bfseries 115} (2015) 012301},
\href{http://arxiv.org/abs/1502.05382}{{\ttfamily arXiv:1502.05382 [nucl-ex]}}.

\bibitem{Aaboud:2016yar}
{ATLAS Collaboration}, {\em {Measurements of long-range azimuthal anisotropies
  and associated Fourier coefficients for $pp$ collisions at $\sqrt{s}=5.02$
  and $13$ TeV and $p$+Pb collisions at $\sqrt{s_{\mathrm{NN}}}=5.02$ TeV with
  the ATLAS detector}},
  \href{http://dx.doi.org/10.1103/PhysRevC.96.024908}{Phys. Rev. C {\bfseries
  96} (2017) 024908},
\href{http://arxiv.org/abs/1609.06213}{{\ttfamily arXiv:1609.06213 [nucl-ex]}}.

\bibitem{Dusling:2013qoz}
K.~Dusling and R.~Venugopalan, {\em {Comparison of the color glass condensate
  to dihadron correlations in proton-proton and proton-nucleus collisions}},
  \href{http://dx.doi.org/10.1103/PhysRevD.87.094034}{Phys. Rev. D {\bfseries
  87} (2013) 094034},
\href{http://arxiv.org/abs/1302.7018}{{\ttfamily arXiv:1302.7018 [hep-ph]}}.

\bibitem{Bozek:2013uha}
P.~Bozek and W.~Broniowski, {\em {Collective dynamics in high-energy
  proton-nucleus collisions}},
  \href{http://dx.doi.org/10.1103/PhysRevC.88.014903}{Phys. Rev. C {\bfseries
  88} (2013) 014903},
\href{http://arxiv.org/abs/1304.3044}{{\ttfamily arXiv:1304.3044 [nucl-th]}}.

\bibitem{Borghini:2000sa}
N.~Borghini, P.~M. Dinh,  and J.-Y. Ollitrault, {\em {New method for measuring
  azimuthal distributions in nucleus-nucleus collisions}},
  \href{http://dx.doi.org/10.1103/PhysRevC.63.054906}{Phys.~Rev.~C {\bfseries
  63} (2001) 054906}, \href{http://arxiv.org/abs/nucl-th/0007063}{{\ttfamily
  arXiv:nucl-th/0007063 [nucl-th]}}.

\bibitem{Bilandzic:2010jr}
A.~Bilandzic, R.~Snellings,  and S.~Voloshin, {\em {Flow analysis with
  cumulants: Direct calculations}},
  \href{http://dx.doi.org/10.1103/PhysRevC.83.044913}{Phys.~Rev.~C {\bfseries
  83} (2011) 044913}, \href{http://arxiv.org/abs/1010.0233}{{\ttfamily
  arXiv:1010.0233 [nucl-ex]}}.

\bibitem{Bilandzic:2013kga}
A.~Bilandzic, C.~H. Christensen, K.~Gulbrandsen, A.~Hansen,  and Y.~Zhou, {\em
  {Generic framework for anisotropic flow analyses with multiparticle azimuthal
  correlations}}, \href{http://dx.doi.org/10.1103/PhysRevC.89.064904}{Phys.
  Rev. C {\bfseries 89} (2014) 064904},
\href{http://arxiv.org/abs/1312.3572}{{\ttfamily arXiv:1312.3572 [nucl-ex]}}.

\bibitem{Jia:2017hbm}
J.~Jia, M.~Zhou,  and A.~Trzupek, {\em {Revealing long-range multiparticle
  collectivity in small collision systems via subevent cumulants}},
  \href{http://dx.doi.org/10.1103/PhysRevC.96.034906}{Phys. Rev. C {\bfseries
  96} (2017) 034906},
\href{http://arxiv.org/abs/1701.03830}{{\ttfamily arXiv:1701.03830 [nucl-th]}}.

\bibitem{Aad:2015gqa}
{ATLAS Collaboration}, {\em {Observation of Long-Range Elliptic Azimuthal
  Anisotropies in $\sqrt{s}=$13 and 2.76 TeV $pp$ Collisions with the ATLAS
  Detector}}, \href{http://dx.doi.org/10.1103/PhysRevLett.116.172301}{Phys.
  Rev. Lett. {\bfseries 116} (2016) 172301},
\href{http://arxiv.org/abs/1509.04776}{{\ttfamily arXiv:1509.04776 [hep-ex]}}.

\bibitem{Khachatryan:2016txc}
{CMS Collaboration}, {\em {Evidence for collectivity in pp collisions at the
  LHC}}, \href{http://dx.doi.org/10.1016/j.physletb.2016.12.009}{Phys. Lett. B
  {\bfseries 765} (2017) 193--220},
\href{http://arxiv.org/abs/1606.06198}{{\ttfamily arXiv:1606.06198 [nucl-ex]}}.

\bibitem{Huo:2017nms}
P.~Huo, K.~Gajdošová, J.~Jia,  and Y.~Zhou, {\em {Importance of non-flow in
  mixed-harmonic multi-particle correlations in small collision systems}},
  \href{http://dx.doi.org/10.1016/j.physletb.2017.12.035}{Phys. Lett. B
  {\bfseries 777} (2018) 201},
\href{http://arxiv.org/abs/1710.07567}{{\ttfamily arXiv:1710.07567 [nucl-ex]}}.

\bibitem{Aaboud:2017acw}
{ATLAS Collaboration}, {\em {Measurement of multi-particle azimuthal
  correlations in $pp$, $p+$Pb and low-multiplicity Pb$+$Pb collisions with the
  ATLAS detector}},
  \href{http://dx.doi.org/10.1140/epjc/s10052-017-4988-1}{Eur. Phys. J. C
  {\bfseries 77} (2017) 428},
\href{http://arxiv.org/abs/1705.04176}{{\ttfamily arXiv:1705.04176 [hep-ex]}}.

\bibitem{Aaboud:2017blb}
{ATLAS Collaboration}, {\em {Measurement of multi-particle azimuthal
  correlations with the subevent cumulant method in $pp$ and $p$+Pb collisions
  with the ATLAS detector at the LHC}},
  \href{http://dx.doi.org/10.1103/PhysRevC.97.024904}{Phys. Rev. C {\bfseries
  97} (2018) 024904},
\href{http://arxiv.org/abs/1708.03559}{{\ttfamily arXiv:1708.03559 [hep-ex]}}.

\bibitem{Sirunyan:2017uyl}
{CMS Collaboration}, {\em {Observation of correlated azimuthal anisotropy
  Fourier harmonics in pp and pPb collisions at the LHC}},
  \href{http://dx.doi.org/10.1103/PhysRevLett.120.092301}{Phys. Rev. Lett.
  {\bfseries 120} (2018) 092301},
\href{http://arxiv.org/abs/1709.09189}{{\ttfamily arXiv:1709.09189 [nucl-ex]}}.

\bibitem{Aad:2008zzm}
{ATLAS Collaboration}, {\em {The ATLAS Experiment at the CERN Large Hadron
  Collider}},
\href{http://dx.doi.org/10.1088/1748-0221/3/08/S08003}{JINST {\bfseries 3}
  (2008) S08003}.

\bibitem{Aad:ID}
{ATLAS Collaboration}, {\em {The ATLAS Inner Detector commissioning and
  calibration}},
  \href{http://dx.doi.org/10.1140/epjc/s10052-010-1366-7}{Eur.~Phys.~J.~C
  {\bfseries 70} (2010) 787}, \href{http://arxiv.org/abs/1004.5293}{{\ttfamily
  arXiv:1004.5293 [physics.ins-det]}}.

\bibitem{ATLAS-TDR-19}
{ATLAS Collaboration}, {\em ATLAS Insertable B-Layer Technical Design Report},
  Atlas-tdr-19, 2010, \url{https://cds.cern.ch/record/1291633}, {\textit{ATLAS
  Insertable B-Layer Technical Design Report Addendum}}, ATLAS-TDR-19-ADD-1,
  2012, {\scriptsize{URL}}: \url{https://cds.cern.ch/record/1451888}.

\bibitem{Aad:2012xs}
{ATLAS Collaboration}, {\em {Performance of the ATLAS Trigger system in 2010}},
  \href{http://dx.doi.org/10.1140/epjc/s10052-011-1849-1}{Eur. Phys. J.~C
  {\bfseries 72} (2012) 1849},
\href{http://arxiv.org/abs/1110.1530}{{\ttfamily arXiv:1110.1530 [hep-ex]}}.

\bibitem{Aaboud:2016leb}
{ATLAS Collaboration}, {\em {Performance of the ATLAS Trigger system in 2015}},
  \href{http://dx.doi.org/10.1140/epjc/s10052-017-4852-3}{Eur. Phys. J. C
  {\bfseries 77} (2017) 317},
\href{http://arxiv.org/abs/1611.09661}{{\ttfamily arXiv:1611.09661 [hep-ex]}}.

\bibitem{Aad:2016mok}
{ATLAS Collaboration}, {\em {Charged-particle distributions in $\sqrt{s}$=13
  TeV $pp$ interactions measured with the ATLAS detector at the LHC}},
  \href{http://dx.doi.org/10.1016/j.physletb.2016.04.050}{Phys. Lett. B
  {\bfseries 758} (2016) 67},
\href{http://arxiv.org/abs/1602.01633}{{\ttfamily arXiv:1602.01633 [hep-ex]}}.

\bibitem{atlas:3}
{ATLAS Collaboration}, {\em {Performance of the ATLAS Minimum Bias and Forward
  Detector Triggers in pPb collisions}}, ATLAS-CONF-2013-104,
  \url{https://cds.cern.ch/record/1624013}.

\bibitem{Sjostrand:2007gs}
T.~{Sj\"ostrand}, S.~Mrenna,  and P.~Z. Skands, {\em {A brief introduction to
  PYTHIA 8.1}}, \href{http://dx.doi.org/10.1016/j.cpc.2008.01.036}{Comput.
  Phys. Commun. {\bfseries 178} (2008) 852},
\href{http://arxiv.org/abs/0710.3820}{{\ttfamily arXiv:0710.3820 [hep-ph]}}.

\bibitem{atlas:6}
{ATLAS Collaboration}, {\em {ATLAS tunes of PYTHIA 6 and Pythia 8 for MC11}},
  ATLAS-PHYS-PUB-2011-009, \url{https://cds.cern.ch/record/1363300}.

\bibitem{Gyulassy:1994ew}
M.~Gyulassy and X.-N. Wang, {\em {HIJING 1.0: A Monte Carlo program for parton
  and particle production in high-energy hadronic and nuclear collisions}},
  \href{http://dx.doi.org/10.1016/0010-4655(94)90057-4}{Comput.~Phys.~Commun.
  {\bfseries 83} (1994) 307},
  \href{http://arxiv.org/abs/nucl-th/9502021}{{\ttfamily
  arXiv:nucl-th/9502021}}.

\bibitem{Agostinelli:2002hh}
{GEANT4} Collaboration, S.~Agostinelli {et~al.}, {\em {GEANT4: A Simulation
  toolkit}},
\href{http://dx.doi.org/10.1016/S0168-9002(03)01368-8}{Nucl.~Instrum.~Meth.~A
  {\bfseries 506} (2003) 250}.

\bibitem{Aad:2010ah}
{ATLAS Collaboration}, {\em {The ATLAS Simulation Infrastructure}},
  \href{http://dx.doi.org/10.1140/epjc/s10052-010-1429-9}{Eur.~Phys.~J.~C
  {\bfseries 70} (2010) 823}, \href{http://arxiv.org/abs/1005.4568}{{\ttfamily
  arXiv:1005.4568 [physics.ins-det]}}.

\bibitem{Aaboud:2016jnr}
{ATLAS Collaboration}, {\em {Measurement of forward-backward multiplicity
  correlations in lead-lead, proton-lead and proton-proton collisions with the
  ATLAS detector}}, \href{http://dx.doi.org/10.1103/PhysRevC.95.064914}{Phys.
  Rev. C {\bfseries 95} (2017) 064914},
\href{http://arxiv.org/abs/1606.08170}{{\ttfamily arXiv:1606.08170 [hep-ex]}}.

\bibitem{DiFrancesco:2016srj}
P.~Di~Francesco, M.~Guilbaud, M.~Luzum,  and J.-Y. Ollitrault, {\em {Systematic
  procedure for analyzing cumulants at any order}},
  \href{http://dx.doi.org/10.1103/PhysRevC.95.044911}{Phys. Rev. C {\bfseries
  95} (2017) 044911},
\href{http://arxiv.org/abs/1612.05634}{{\ttfamily arXiv:1612.05634 [nucl-th]}}.

\bibitem{Giacalone:2016afq}
G.~Giacalone, L.~Yan, J.~Noronha-Hostler,  and J.-Y. Ollitrault, {\em
  {Symmetric cumulants and event-plane correlations in Pb + Pb collisions}},
  \href{http://dx.doi.org/10.1103/PhysRevC.94.014906}{Phys. Rev. C {\bfseries
  94} (2016) 014906},
\href{http://arxiv.org/abs/1605.08303}{{\ttfamily arXiv:1605.08303 [nucl-th]}}.

\bibitem{Das:2017ned}
S.~J. Das, G.~Giacalone, P.-A. Monard,  and J.-Y. Ollitrault, {\em {Relating
  centrality to impact parameter in nucleus-nucleus collisions}},
  \href{http://dx.doi.org/10.1103/PhysRevC.97.014905}{Phys. Rev. C {\bfseries
  97} (2018) 014905},
\href{http://arxiv.org/abs/1708.00081}{{\ttfamily arXiv:1708.00081 [nucl-th]}}.

\bibitem{Bhalerao:2013ina}
R.~S. Bhalerao, J.-Y. Ollitrault,  and S.~Pal, {\em {Event-plane correlators}},
  \href{http://dx.doi.org/10.1103/PhysRevC.88.024909}{Phys. Rev. C {\bfseries
  88} (2013) 024909},
\href{http://arxiv.org/abs/1307.0980}{{\ttfamily arXiv:1307.0980 [nucl-th]}}.

\bibitem{Aad:2012bu}
{ATLAS Collaboration}, {\em {Measurement of the azimuthal anisotropy for
  charged particle production in $\sqrt{s_{_{NN}}}=2.76$ TeV lead-lead
  collisions with the ATLAS detector}},
  \href{http://dx.doi.org/10.1103/PhysRevC.86.014907}{Phys.~Rev.~C {\bfseries
  86} (2012) 014907}, \href{http://arxiv.org/abs/1203.3087}{{\ttfamily
  arXiv:1203.3087 [hep-ex]}}.

\bibitem{Bozek:2010vz}
P.~Bozek, W.~Broniowski,  and J.~Moreira, {\em {Torqued fireballs in
  relativistic heavy-ion collisions}},
  \href{http://dx.doi.org/10.1103/PhysRevC.83.034911}{Phys.~Rev.~C {\bfseries
  83} (2011) 034911}, \href{http://arxiv.org/abs/1011.3354}{{\ttfamily
  arXiv:1011.3354 [nucl-th]}}.

\bibitem{Pang:2014pxa}
L.-G. Pang, G.-Y. Qin, V.~Roy, X.-N. Wang,  and G.-L. Ma, {\em {Longitudinal
  decorrelation of anisotropic flows in heavy-ion collisions at the CERN Large
  Hadron Collider}},
  \href{http://dx.doi.org/10.1103/PhysRevC.91.044904}{Phys.~Rev.~C {\bfseries
  91} (2015) 044904}, \href{http://arxiv.org/abs/1410.8690}{{\ttfamily
  arXiv:1410.8690 [nucl-th]}}.

\bibitem{Khachatryan:2015oea}
{CMS Collaboration}, {\em {Evidence for transverse momentum and pseudorapidity
  dependent event plane fluctuations in PbPb and pPb collisions}},
  \href{http://dx.doi.org/10.1103/PhysRevC.92.034911}{Phys. Rev. C {\bfseries
  92} (2015) 034911},
\href{http://arxiv.org/abs/1503.01692}{{\ttfamily arXiv:1503.01692 [nucl-ex]}}.

\bibitem{Aaboud:2017tql}
{ATLAS Collaboration}, {\em {Measurement of longitudinal flow de-correlations
  in Pb+Pb collisions at $\sqrt{s_\mathrm{NN}} = 2.76$ and 5.02 TeV with the
  ATLAS detector}},
  \href{http://dx.doi.org/10.1140/epjc/s10052-018-5605-7}{Eur. Phys. J. C
  {\bfseries 78} (2018) 142},
\href{http://arxiv.org/abs/1709.02301}{{\ttfamily arXiv:1709.02301 [nucl-ex]}}.

\bibitem{Aad:2014fla}
{ATLAS Collaboration}, {\em {Measurement of event-plane correlations in
  $\sqrt{s_{NN}}=2.76$ TeV lead-lead collisions with the ATLAS detector}},
  \href{http://dx.doi.org/10.1103/PhysRevC.90.024905}{Phys.~Rev.~C {\bfseries
  90} (2014) 024905}, \href{http://arxiv.org/abs/1403.0489}{{\ttfamily
  arXiv:1403.0489 [hep-ex]}}.

\bibitem{Aad:2015lwa}
{ATLAS Collaboration}, {\em {Measurement of the correlation between flow
  harmonics of different order in lead-lead collisions at $\sqrt{s_{NN}}$=2.76
  TeV with the ATLAS detector}},
  \href{http://dx.doi.org/10.1103/PhysRevC.92.034903}{Phys. Rev. C {\bfseries
  92} (2015) 034903},
\href{http://arxiv.org/abs/1504.01289}{{\ttfamily arXiv:1504.01289 [hep-ex]}}.

\bibitem{ALICE:2016kpq}
{ALICE} Collaboration, J.~Adam {et~al.}, {\em {Correlated Event-by-Event
  Fluctuations of Flow Harmonics in Pb+Pb Collisions at $\sqrt{s_{NN}}=2.76$
  TeV}}, \href{http://dx.doi.org/10.1103/PhysRevLett.117.182301}{Phys. Rev.
  Lett. {\bfseries 117} (2016) 182301},
\href{http://arxiv.org/abs/1604.07663}{{\ttfamily arXiv:1604.07663 [nucl-ex]}}.

\bibitem{ATL-GEN-PUB-2016-002}
{ATLAS Collaboration}, {\em {ATLAS Computing Acknowledgements}},
  {ATL-GEN-PUB-2016-002}, \url{https://cds.cern.ch/record/2202407}.

\end{thebibliography}\endgroup
\clearpage 
 
\begin{flushleft}
{\Large The ATLAS Collaboration}

\bigskip

M.~Aaboud$^\textrm{\scriptsize 34d}$,    
G.~Aad$^\textrm{\scriptsize 99}$,    
B.~Abbott$^\textrm{\scriptsize 125}$,    
O.~Abdinov$^\textrm{\scriptsize 13,*}$,    
B.~Abeloos$^\textrm{\scriptsize 129}$,    
D.K.~Abhayasinghe$^\textrm{\scriptsize 91}$,    
S.H.~Abidi$^\textrm{\scriptsize 164}$,    
O.S.~AbouZeid$^\textrm{\scriptsize 39}$,    
N.L.~Abraham$^\textrm{\scriptsize 153}$,    
H.~Abramowicz$^\textrm{\scriptsize 158}$
% ...
\bigskip
\\

% ...
$^{13}$Institute of Physics, Azerbaijan Academy of Sciences, Baku; Azerbaijan.\\
% ...
$^{34}$$^{(a)}$Facult\'e des Sciences Ain Chock, R\'eseau Universitaire de Physique des Hautes Energies - Universit\'e Hassan II, Casablanca;$^{(b)}$Centre National de l'Energie des Sciences Techniques Nucleaires (CNESTEN), Rabat;$^{(c)}$Facult\'e des Sciences Semlalia, Universit\'e Cadi Ayyad, LPHEA-Marrakech;$^{(d)}$Facult\'e des Sciences, Universit\'e Mohamed Premier and LPTPM, Oujda;$^{(e)}$Facult\'e des sciences, Universit\'e Mohammed V, Rabat; Morocco.\\
% ...
$^{39}$Niels Bohr Institute, University of Copenhagen, Copenhagen; Denmark.\\
% ...
$^{91}$Department of Physics, Royal Holloway University of London, Egham; United Kingdom.\\
% ...
$^{99}$CPPM, Aix-Marseille Universit\'e, CNRS/IN2P3, Marseille; France.\\
% ...
$^{125}$Homer L. Dodge Department of Physics and Astronomy, University of Oklahoma, Norman OK; United States of America.\\
% ...
$^{129}$LAL, Universit\'e Paris-Sud, CNRS/IN2P3, Universit\'e Paris-Saclay, Orsay; France.\\
% ...
$^{153}$Department of Physics and Astronomy, University of Sussex, Brighton; United Kingdom.\\
% ...
$^{158}$Raymond and Beverly Sackler School of Physics and Astronomy, Tel Aviv University, Tel Aviv; Israel.\\
% ...
$^{164}$Department of Physics, University of Toronto, Toronto ON; Canada.\\
% ...
$^{a}$ Also at Borough of Manhattan Community College, City University of New York, NY; United States of America.\\
$^{b}$ Also at California State University, East Bay; United States of America.\\
$^{c}$ Also at Centre for High Performance Computing, CSIR Campus, Rosebank, Cape Town; South Africa.\\
$^{d}$ Also at CERN, Geneva; Switzerland.\\
$^{e}$ Also at CPPM, Aix-Marseille Universit\'e, CNRS/IN2P3, Marseille; France.\\
% ...
$^{*}$ Deceased

\end{flushleft}


\end{document}